\documentclass[aip,apl,reprint,twocolumn]{revtex4-1}
\usepackage{graphicx}
\usepackage{dcolumn}
\usepackage{bm}
\usepackage{amssymb,amsmath}
\usepackage{gensymb}
\bibstyle{apsrev}
\usepackage{hyperref}
\usepackage{color}
\usepackage[normalem]{ulem}

\begin{document}
\preprint{PREPRINT}

\title[Molecules on rails: friction anisotropy and preferential sliding directions of organic nanocrystallites on two-dimensional materials]{Molecules on rails: friction anisotropy and preferential sliding directions of organic nanocrystallites on two-dimensional materials}

\author{Borislav Vasi\'c}\email{bvasic@ipb.ac.rs}
\affiliation{Graphene Laboratory of Center for Solid State Physics and New Materials, Institute of Physics Belgrade, University of Belgrade, Pregrevica 118, 11080 Belgrade, Serbia}
\author{Igor Stankovi\'c}\email{igor.stankovic@ipb.ac.rs}
\affiliation{Scientific Computing Laboratory, Center for the Study of Complex Systems, Institute of Physics Belgrade, University of Belgrade, Pregrevica 118, 11080 Belgrade, Serbia.}
\author{Aleksandar Matkovi\'c}\email{aleksandar.matkovic@unileoben.ac.at}
\affiliation{Institute of Physics, Montanuniversit{\"a}t Leoben, Franz Josef Strasse 18, 8700 Leoben, Austria}
\author{Markus Kratzer}
\affiliation{Institute of Physics, Montanuniversit{\"a}t Leoben, Franz Josef Strasse 18, 8700 Leoben, Austria}
\author{Christian Ganser}
\affiliation{Institute of Physics, Montanuniversit{\"a}t Leoben, Franz Josef Strasse 18, 8700 Leoben, Austria, {\itshape Present address:} Department of Physics, Nagoya University, Furo-cho, Chikusa-ku, Nagoya, Aichi, 464-8602, Japan}
\author{Rado\v s Gaji\'c}
\affiliation{Graphene Laboratory of Center for Solid State Physics and New Materials, Institute of Physics Belgrade, University of Belgrade, Pregrevica 118, 11080 Belgrade, Serbia}
\author{Christian Teichert}
\affiliation{Institute of Physics, Montanuniversit{\"a}t Leoben, Franz Josef Strasse 18, 8700 Leoben, Austria}

\date{\today}

\begin{abstract}
Two-dimensional (2D) materials are envisaged as ultra-thin solid lubricants for nano-mechanical systems. So far, their frictional properties at the nanoscale have been studied by standard friction force microscopy. However, lateral manipulation of nanoparticles is a more suitable method to study the dependence of friction on the crystallography of two contacting surfaces. Still, such experiments are lacking. In this study, we combine atomic force microscopy (AFM) based lateral manipulation and molecular dynamics simulations in order to investigate the movements of organic needle-like nanocrystallites grown by van der Waals epitaxy on graphene and hexagonal boron nitride. We observe that nanoneedle fragments -- when pushed by an AFM tip -- do not move along the original pushing directions. Instead, they slide on the 2D materials preferentially along the needles' growth directions, which act as invisible rails along commensurate directions. Further, when the nanocrystallites were rotated by applying a torque with the AFM tip across the preferential sliding directions, we find an increase of the torsional signal of the AFM cantilever. We demonstrate in conjunction with simulations that both, the significant friction anisotropy and preferential sliding directions are determined by the complex epitaxial relation and arise from the commensurate and incommensurate states between the organic nanocrystallites and the 2D materials.
\end{abstract}

\maketitle

\section{Introduction}

Bulk layered materials such as graphite, transition-metal dichalcogenides, and hexagonal boron-nitride exhibit low friction because of their lamelar structure and easy shearing of layers. For these reasons, they are widely used as solid lubricants \cite{Erdemir_review}. Still, bulky lubricants are not appropriate for nanodevices where ultra-thin coatings with a maximal thickness of only several nanometers are required \cite{friction_Carpick}. As a result, atomically thin, two-dimensional (2D) materials and especially graphene (Gr) have been recently envisaged as solid lubricants for friction and wear reduction in nano-mechanical systems \cite{graphene_lubricant, friction_Bennewitz, Lee_ACSnano, friction_reduction_Moseler, Berman_adv_func_mat, Vasic_carbon_2017}.

Layered materials are single crystals with van der Waals bonding in only one direction, allowing exposure of atomically flat and dangling-bond free surfaces by simple mechanical cleavage. Therefore, besides the technological applications, they are also suitable for fundamental tribological studies mostly performed by atomic force microscopy (AFM) \cite{Sheehan_science_1996, Falvo_prb_2000, Tranvouez_nanotechnology_2009, Bennewitz_prb_2014, Sheehan_nanolett, Dienwiebel_prl_2004, Schirmeisen_friction_duality, Schirmeisen_scaling_laws, duerig_science_2015, Meyer_science_2016, Baykara_nat_comm}. These studies demonstrated that the substrates' crystal structure determines several fundamental properties, like the existence of friction anisotropy \cite{Sheehan_science_1996, Falvo_prb_2000, Tranvouez_nanotechnology_2009, Bennewitz_prb_2014}, preferential sliding directions \cite{Sheehan_science_1996, Sheehan_nanolett}, and structural lubricity, a state with a low friction between two surfaces sliding through incommensurate states \cite{Hirano_Shinjo_prl_1991, Dienwiebel_prl_2004, Schirmeisen_friction_duality, Schirmeisen_scaling_laws, prl_microscale_superlubricity, deWijn_prb_2012, Salmeron_acs_nano, zheng_sci_rep_2014, duerig_science_2015, berman_science_2015, Meyer_science_2016, Baykara_nat_comm}. Still, the influence of the epitaxial relation between two contacting surfaces on the resulting sliding directions and friction anisotropy has been explored much less. Until now, the underlying epitaxial relations were considered only for simple triangular and square lattices \cite{Sheehan_science_1996, Sheehan_nanolett, deWijn_prb_2012}.

Frictional properties of 2D materials were investigated so far only by AFM derived friction force microscopy (FFM) \cite{friction_Carpick, friction_Bennewitz, Lee_ACSnano, friction_reduction_Moseler, Berman_adv_func_mat, Vasic_carbon_2017, friction_Lee, friction_Park, friction_Meyer, Filleter_GO}. However, the often ill-defined structure of the AFM tip is an obstacle to study friction as a function of the relative orientation between the crystal lattices of two contacting surfaces \cite{Sheehan_science_1996, Schirmeisen_jap_2007}. For this purpose, AFM based lateral manipulation \cite{Sheehan_science_1996, Schirmeisen_friction_duality, Schirmeisen_scaling_laws, Baykara_nat_comm, Tranvouez_nanotechnology_2009, Sheehan_nanolett} of particles with well defined crystallographic structures and epitaxial relations to 2D materials is a more appropriate technique than standard FFM.

Van der Walls (vdW) heterostructures consisting of epitaxially grown organic crystallites on 2D materials can serve as an excellent paradigmatic system to explore the influence of the inherent epitaxial relation on the friction during AFM based lateral manipulation. 2D materials are superior substrates for the epitaxial growth \cite{Koma} of organic molecules \cite{Kratzer_review, Teichert_nanolett, Matkovic_scirep_2016, Kim_apl_mater, Kim_adv_mater, Wang_prl, Hersam_nanolett}. They are atomically smooth with no dangling bonds and trapped charges at the interface, thus providing a pure vdW interface between two contacting surfaces. While friction studies are usually constrained by contaminant molecules \cite{Muser_science_1999, Schirmeisen_friction_duality, Baykara_nat_comm} and chemical interactions \cite{Schirmeisen_chemical_interactions}, 2D materials may provide a clean interface between the contacting surfaces. At the same time, organic crystallites form complex epitaxial relations with 2D materials \cite{Kratzer_review, Teichert_nanolett, Matkovic_scirep_2016}, while their strong intrinsic anisotropy makes them suitable for AFM studies of friction anisotropy and related phenomena \cite{fr_anisotropy_organic_prl_1994, carpick_trib_lett_1999, fr_anisotropy_organic_prl_2010a, fr_anisotropy_organic_prl_2010b, ocal_nanoscale2017}.

In this work, we consider, as representative vdW heterostructures, organic, needle-like nanocrystallites (also called nanoneedles, nanowires, or nanorods) formed by para-hexaphenyl (6P) molecules grown by vdW epitaxy on Gr and hexagonal boron nitride (hBN). These organic nanocrystallites are large enough to be considered as bulk structures, they are strongly anisotropic and stable under ambient conditions. By combined AFM manipulations and molecular dynamics (MD) simulations, we investigate lateral movements of 6P needles on 2D materials. We identified preferential sliding directions, \textit{i.e.}, registry states, which are different from the pushing directions defined by the AFM tip movement. During rotations of 6P needles, an increased friction force was observed when crossing the registry states on the 2D substrates, indicating a pronounced friction anisotropy.

\section{Experimental}

\subsection{Sample preparation}

Flakes of single- and multi-layer Gr and multi-layer hBN -- prepared by mechanical exfoliation and transferred onto SiO$_2$/Si following known recipes \cite{novoselov_science2004} -- have been used as substrates for the growth of parahexaphenyl (6P). The molecules were deposited by hot wall epitaxy (HWE)~\cite{lopez1978hot}. As a source material, commercially available 6P from TCI Chemicals (S0220) was used. The base pressure of the HWE chamber was $\sim$2$\times$10$^{-6}$~mbar, source and wall temperatures were kept fixed at 510~K and 520~K, respectively. Substrate temperature during the growth was varied between 380~K and 420~K. The amount of 6P deposited on the surface of the samples corresponds to an equivalent of 0.8-1.2 monolayers of 6P. Here, a monolayer is defined by the molecular density in the beta-phase 6P (001) plane (4.4$\times$10$^{14}$ molecules/$\mathrm{cm}^2$) \cite{potocar_prb}. On both, Gr and hBN, 6P molecules were found to form three-dimensional needle-like crystallites~\cite{Teichert_nanolett, balzer2013Gr6Porientation, kratzer2013temperature, Kratzer_review, Matkovic_scirep_2016}. In the case of 6P needles, not always the molecules assume a "lying" orientation having their long molecular axes (LMA) parallel to the substrate plane~\cite{Kratzer_review, hlawacek2013nucleation, simbrunner2013epitaxial}. These needle-like crystallites are large enough to be considered as $\beta$-phase bulk 6P, in which the molecules have a herringbone motif~\cite{baker1993crystal}. The chosen growth parameters result in tens of micrometer long and 5-10 nm tall 6P needles that follow six directions dictated by the epitaxial relation between 6P and the 2D material substrate~\cite{Kratzer_review, Matkovic_scirep_2016, simbrunner2013epitaxial}.\par

\subsection{AFM measurements}

AFM measurements were performed using an NTEGRA Prima AFM system from NT-MDT and an Asylum Research MFP 3D device. AFM imaging and manipulations were done with NSG01 (Gr substrate) and FMG01 (hBN substrate) probes from NT-MDT. Spring constant calibration of AFM cantilevers was performed via the thermal noise method \cite{Hutter_calibration}, employing the MFP 3D AFM. All measurements were performed under ambient conditions.

After initial sample imaging in tapping mode, the first step was to prepare a short 6P needle suitable for AFM manipulations. For this purpose, an appropriate long 6P needle was selected and then cut by AFM manipulation in contact mode \cite{Bogild_small_2006}. The typical procedure is illustrated in Fig. S4 of ESI. Cutting was repeated if needed for several times until a short needle of around $200 \ \mathrm{nm} - 400 \ \mathrm{nm}$ was obtained.

AFM manipulations were done in a standard way following procedures in Refs. \cite{Samuelson_APL, Lindelof_Nanotechnology, Schirmeisen_jap_2007}. A selected short needle was first imaged in tapping mode. Then we switched to contact mode. The AFM probe was moved in x-direction with the cantilever's long axis oriented in y-direction like conventionally done in friction force microscopy. The AFM tip was pushed towards one of the needle's endings for a certain distance. The reason we pushed needles from their endings was because we were not interested in the trivial case where needles, pushed in the center were just translated along the tip path direction. The path length was in the range of $500 \ \mathrm{nm}- 1500 \ \mathrm{nm}$, while the normal force (determined by the AFM cantilever bending) during the pushing was around $100 \ \mathrm{nN}$. After each manipulation step, the needle was imaged in tapping mode in order to visualize its movement. This procedure was repeated by around 100 times with the same probe, and it was performed for selected short needles on both, Gr and hBN. Compared to AFM manipulation experiments of nanorods \cite{Gnecco2010}, here all movements were performed just once, along a single line, while the focus was on the influence of the crystal structure of substrates on the resulting motion.

In each manipulation step, simultaneously with movements of 6P needles, the lateral force - proportional to the AFM cantilever torsion - was recorded. The lateral force signal was calibrated according to the procedure introduced by Varenberg \textit{et al.} \cite{Varenberg}. All AFM manipulations presented in the paper were done along the x-axis. In cases where needles were almost aligned with the x-axis, they were pushed along the y-axis to reorient them. However, these manipulation steps were not taken into consideration since lateral forces could not be measured.

\subsection{Molecular dynamics simulations}

In our atomistic model, a $90$~\r{A}~$\times \ 300$~\r{A} 6P needle was placed on a $380$~\r{A}~$\times \ 380$~\r{A} Gr sheet. Periodic boundary conditions were set in $x$ and $y$ direction. The crystallographic data for the unit cell of $\beta$-phase 6P bulk was taken from the paper of Baker et al.~\cite{baker1993crystal}. The lattice parameters of the monoclinic unit cell including two molecules were $a~=~26.241$~\r{A}, $b~=~5.568$~\r{A}, and $c~=~8.091$~\r{A} and the angle $\beta~=~98.17^\circ$. The herringbone arrangement of the unit cell was defined by the intersection angles $\omega~=~26^\circ$ and $\phi~=~71^\circ$, and setting angle $\Theta~=~55^\circ$. The herringbone angle, calculated from previous values, was $\tau~=~61^\circ$. The contact plane of 6P needle was $(11\overline{1})$~\cite{Teichert_nanolett, balzer2013Gr6Porientation}.

The interatomic forces within Gr were derived using the appropriate Tersoff potential~\cite{Tersoff_CC}. Interactions between 6P molecules were modeled using empirical CHARMM force field parameters~\cite{CHARMM}. The adhesion forces between the carbon atoms in 6P molecules and Gr were modeled with a registry dependent Kolmogorov-Crespi potential~\cite{PhysRevB.71.235415}. For the interaction of C atoms in Gr with hydrogen atoms of the 6P molecules, CHARMM force field parameters were utilized.

The molecular dynamics (MD) simulations were performed using LAMMPS, a commonly used distributed classical MD code~\cite{LAMMPS}. The $15$~\r{A} thick 6P needle was displaced on the Gr sheet with steps of 0.5~fs. The top-most layer of 6P molecules had relative position fixed, while the following three layers towards the interface with Gr and the Gr substrate itself were thermalized at $300$~K. The top layer of the molecules was used to move the needle on the Gr surface. The initial configuration was equilibrated for 1~ns. The distance between Gr and the bottom 6P molecules was roughly $3.2$~\r{A}.


\section{Results and Discussion}

The results are presented in five sections. The epitaxial relations between 6P molecules and hBN/Gr are elaborated in the first part. Then, in the second section, we summarize all experimental results for AFM manipulations of 6P needles. After that, in the third section we analyze the rotation of the needles and the observed friction anisotropy, while the corresponding results of MD simulations are discussed in the fourth section. Finally, in the fifth part, translations of the needles and their preferential sliding directions are discussed.

\subsection{Epitaxial relations}

\begin{figure*}
\centerline{\includegraphics[width=14cm]{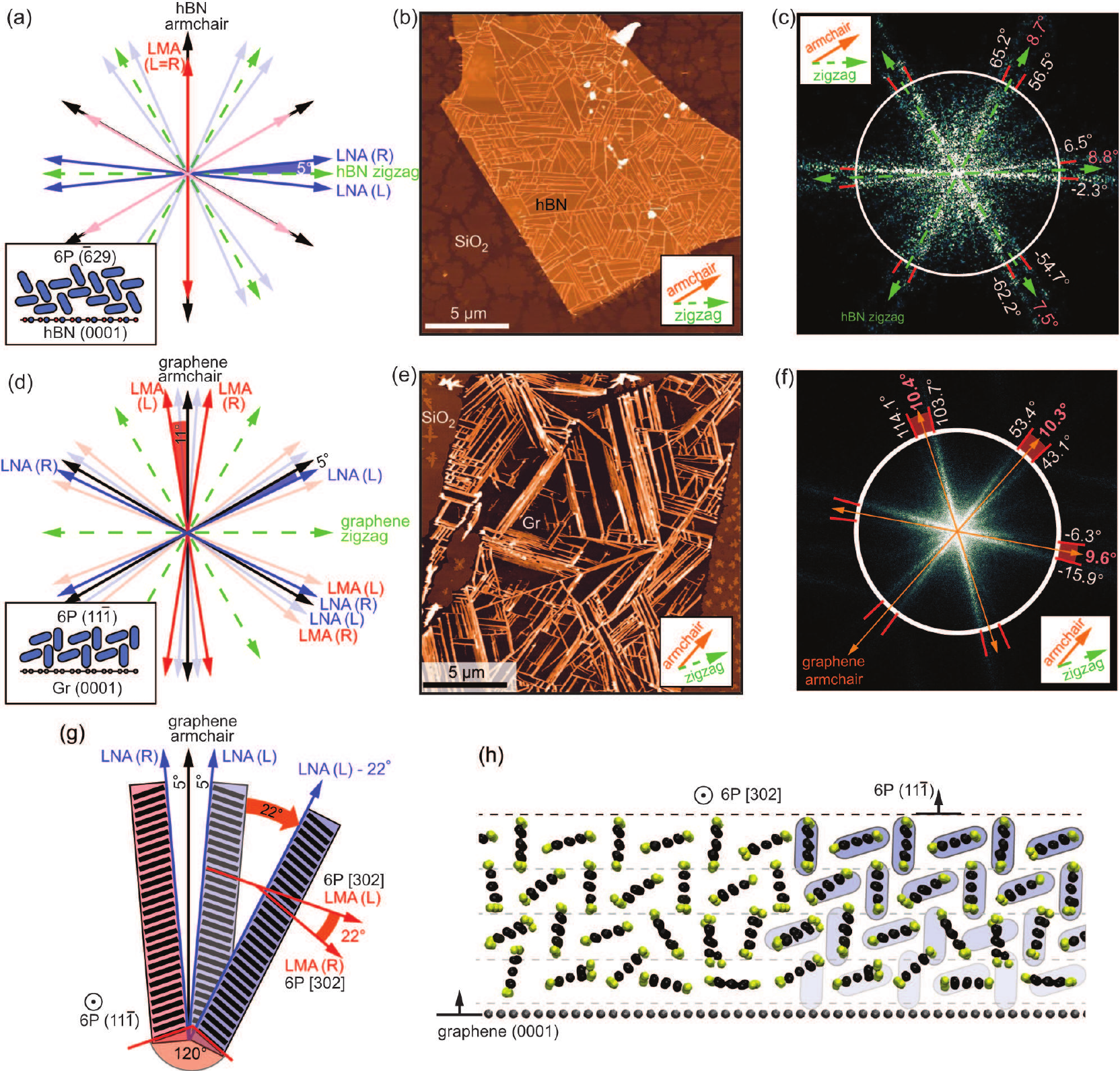}}
\caption{Preferential growth directions of 6P on hBN and Gr: (a), (d) Sketches of the preferential molecule orientation (LMA) and needle growth directions (LNA) with respect to high-symmetry directions (zigzag and armchair) of hBN and Gr, respectively. Insets of (a) and (d) illustrate side views of the molecular packing at the interface with hBN and Gr, respectively, considering only the epitaxial relation between 6P molecules and these two materials. (b), (e) Characteristic AFM images of needles grown on hBN (z-scale is $60\ \mathrm{nm}$) and Gr (z-scale is $20\ \mathrm{nm}$), respectively. Islands observed at the edges of the AFM images are located on SiO$_2$ support and are formed by up-right standing molecules, which is a characteristic growth mode on SiO$_2$. (c), (f) 2D-FFT images of the topographic images (given in (b) and (e)) for hBN and Gr, respectively. The 2D-FFT diagrams are rotated by 90$^\circ$ to match directly with the orientations of the preferential needle growth directions (LNA). 2D-FFT images are generated from binary masks of the topography images set to highlight only the needles. Dashed (green) and solid (orange) arrows, respectively, indicate zigzag and armchair directions of both, Gr and hBN with respect to the $x$-axis of the AFM scanner. White circles are guides to the eye and their radius is $10 \ \mathrm{\mu m^{-1}}$. (g) Schematic illustration, how the 6P needle -- aligned along LNA(L) direction -- falls into a rotationally commensurate registry state when rotated clockwise by $22^\circ$. The resulting state does not coincide with any of LNAs (is not a growth direction). In (a), (d), and (g) solid red arrows indicate LMA directions, solid blue arrows indicate LNA directions, dashed green and solid black arrows indicate zigzag and armchair directions of the 2D material substrate, while "L" and "R" stand for left- and right-handed chiral pairs of the crystallites. More details on epitaxial relations are given in Figs. S1 and S2 of ESI. (h) A side view along LMA of a $300 \ \mathrm{K}$ MD simulation snapshot for a 6P needle on Gr. The overlay in the right part with an ideal molecular packing from the inset in (d) illustrates the (11$\bar{1}$) contact plane of bulk 6P. Results for the top and bottom views of the needles are given in Fig. S3 of ESI.}
\label{structure_fig}
\end{figure*}

Friction anisotropy and preferential sliding directions of 6P needles on 2D material substrates stem from their epitaxial relations. Both individual 6P molecules and 6P needles are intrinsically anisotropic structures and can be considered as quasi one-dimensional objects. As such, there are two main directions to be considered within 6P needles: 1. the long molecular axis (LMA) or the axis along the phenylene backbone of the individual molecules, and 2. the long needle axis (LNA) that indicates the preferred growth direction of the needle on a given substrate \cite{simbrunner2013epitaxial}. Additional data on the orientations of LMA and LNA on Gr are given in Fig. S3 of ESI. Furthermore, preferential growth directions are also influenced by the interactions with the substrate, since the individual molecules tend to adsorb only at specific sites on the substrate. The growth directions of the needles (LNA) are then finally defined by the relation between the LMA and the high-symmetry directions of the substrate (armchair and zigzag directions of Gr and hBN, respectively) and the particular contact plane of the molecular crystal that is best matching the arrangement of the molecules at the interface with the substrate to that of the bulk structure.

If assumed that the molecular crystal remains in the bulk to the very interface, then there is no distinctive registry between the substrate lattice and the deposited lattice, resulting in translational incommensurism \cite{Hooks2001, Mannsfeld2005, Haber2008, Raimondo2011, Raimondo2013, Campione2006}. However, molecular crystals can accommodate large strain, and molecules at the surface frequently rearrange to accommodate both intermolecular interactions that drive the formation of the bulk molecular crystal and interaction with the substrate. As a consequence, the bulk structure of the molecular crystal is not kept at the very interface, and commonly only rotational commensurism is maintained, regardless of the lattice mismatch \cite{Koma}. More details on the epitaxial relation between 6P and Gr/hBN is given in the first section of ESI.

In the case of hBN supported 6P, individual molecules tend to align their LMA exactly with an armchair direction, thus giving the molecular arrangement at the surface well matching the ($\bar{6}$29) plane of bulk 6P~\cite{Matkovic_scirep_2016}. As a result, 6P needles on hBN follow six preferential growth directions as shown in Fig.~\ref{structure_fig}(a). In this case, the orientation of the LNAs are split by $\pm 4.5^\circ$ from a zigzag direction of hBN. The preferential growth directions of 6P needles can be determined from AFM topographic images. A typical topographic image of 6P needles grown on hBN is given in Fig. \ref{structure_fig}(b), while the corresponding 2D fast Fourier transform (2D-FFT) is represented in Fig. \ref{structure_fig}(c). Please note that the 2D-FFT image is rotated by 90$^\circ$ in order to match the real space directions. The bright lines in Fig.~\ref{structure_fig}(c) indicate the preferred growth directions of the needles (LNAs), determined from 2D-FFT with a precision of $\pm$2$^\circ$. The bright lines appear in pairs which are separated from each other by $60^\circ$ due to the sixfold symmetry of hBN. Two bright lines within a single pair are separated from each other by around $9^\circ$, whereas the hBN zigzag directions run along the angle bisector between them. These orientations match quite well the previous observation that the LNA directions split by $\pm 5^\circ$ (with a tolerance of $2^\circ$) from a zigzag direction \cite{Matkovic_scirep_2016}.

For 6P on Gr, preferential growth directions (LNAs) and the orientation of the individual molecules (LMAs) with respect to Gr's high symmetry directions are shown in Fig. \ref{structure_fig}(d). In this case, it has been reported earlier that 6P molecules align with their LMA $\pm$11$^\circ$ rotated from an armchair direction of Gr (graphite) \cite{Teichert_nanolett, balzer2013Gr6Porientation}. The packing motif at the surface then closely resembles the (11$\bar{1}$) plane of bulk 6P \cite{Teichert_nanolett}, thus resulting in a total of six LNA directions split by $\pm 5^\circ$ also from an armchair direction \cite{balzer2013Gr6Porientation, Kratzer_review}. Fig. \ref{structure_fig}(e) depicts a characteristic AFM topography image of the 6P needles on Gr. The corresponding 2D-FFT is given in Fig. \ref{structure_fig}(f). As in the case with hBN substrate, the bright lines in Fig.~\ref{structure_fig}(f) mark the preferred growth directions of the needles. They again appear in pairs which are separated from each other by $60^\circ$ due to the sixfold symmetry of Gr. Now, two bright lines within a single pair are separated from each other by around $10^\circ$, whereas the Gr armchair directions run along the angle bisector between them. These bright lines match very well the prediction that the LNA directions are split by $\pm 5^\circ$ from an armchair direction \cite{balzer2013Gr6Porientation, Kratzer_review}.

Since the LMA of 6P on Gr do not coincide with high symmetry directions of the substrate, it is possible to access only rotationally commensurate states. In the true commensurate states (growth directions), the molecules in the contact with Gr have both, their positions and their LMA matching the preferred adsorption sites of the individual molecules. On the other hand, in a rotationally commensurate state, only the relative angle between 6P LMA and Gr is maintained, while the exact positions (translational symmetry) of the molecules do not match the preferred adsorption sites. Therefore, the crystallites will not grow in these directions. Figure Fig.~\ref{structure_fig}(g) illustrates such a case, and the impact of these states on the friction anisotropy of 6P on Gr will be discussed later.

MD simulations give a realistic picture of the orientation of 6P molecules within a needle and their contact with the substrate. The side view of the MD simulation setup for a 6P needle on Gr is depicted in Fig. \ref{structure_fig}(h) by a snapshot of the MD simulation. The 6P molecules in the top layer of a 4 layer thick needle are fixed to fit the 6P (11$\bar{1}$) plane, while the rest of the system is free to move. 6P molecules from the bottom layer at the interface tend to occupy commensurate states with the underlying Gr with their LMA rotated from an armchair direction by $\pm$11$^\circ$. As a result, the bottom layer consists of almost "flat-lying" 6P molecules which are nearly commensurate with Gr, and "edge-on" molecules, which tend to have the plane of their $\pi$-system normal or inclined to the Gr plane. The bulk herringbone structure (shown as the overlay in Fig. \ref{structure_fig}(h)) consists of molecules with alternate inclination of the short molecular axes of $21.3^\circ$ and $90^\circ$ relative to the substrate. As a result, 6P molecules inside the needle are relaxed as represented by the transition from the bottom layer in contact with Gr to bulk herringbone structure with $(11\overline{1})$ contact plane on the top. Additional data on the MD simulation setup with top and bottom views as well, are presented in Fig. S3 of ESI.

\subsection{AFM manipulations}

\begin{figure}
\centerline{\includegraphics[width=8.3cm]{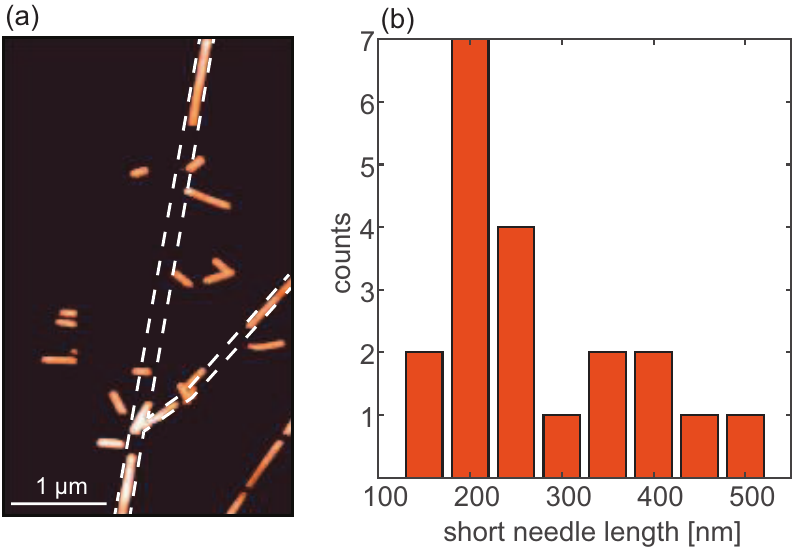}}
\caption{(a) AFM topography image of short needles cut from two former long needles marked by dashed lines. z-scale is $10 \ \mathrm{nm}$. (b) Histogram of the length distribution of the short needles after the cutting.}
\label{needle_length}
\end{figure}

After the growth of 6P needles, AFM in contact mode was employed under ambient conditions to cut them in order to fabricate short needle fragments appropriate for AFM manipulations. The typical procedure for the cutting is illustrated in Fig. S4 of ESI. The AFM topography image in Fig. \ref{needle_length}(a) displays characteristic short needles cut from two long needles. The former edges of these as-grown needles are indicated by dashed lines. The cutting of long needles was a sudden process initiated by a high enough normal load, and we did not observe a significant needle bending prior to the cutting. This is in accordance with the results for manipulations of organic nanofibers \cite{Bogild_small_2006}, but different to InAs nanorods, which were first bent during the AFM manipulation, and then cut \cite{Conache2009}. The histogram of the needle length distribution is presented in Fig. \ref{needle_length}(b) revealing that the typical length of a short needle is around $200 \ \mathrm{nm}$. Beyond this approximate length limit, the cutting was not possible anymore and intended AFM manipulations led only to needle movements which are investigated in detail in the following.

After cutting, the same short needle was pushed by the AFM tip in contact mode for about 100 times. Topographic images were recorded in tapping mode after each manipulation step. The short needles were always pushed from one of their endings and always along the x-axis. This procedure was performed on both, hBN and Gr substrates. Sequences of all AFM tapping mode images are presented in ESI (supplementary movie 1 and 2).

\begin{figure}
\centerline{\includegraphics[width=8.3cm]{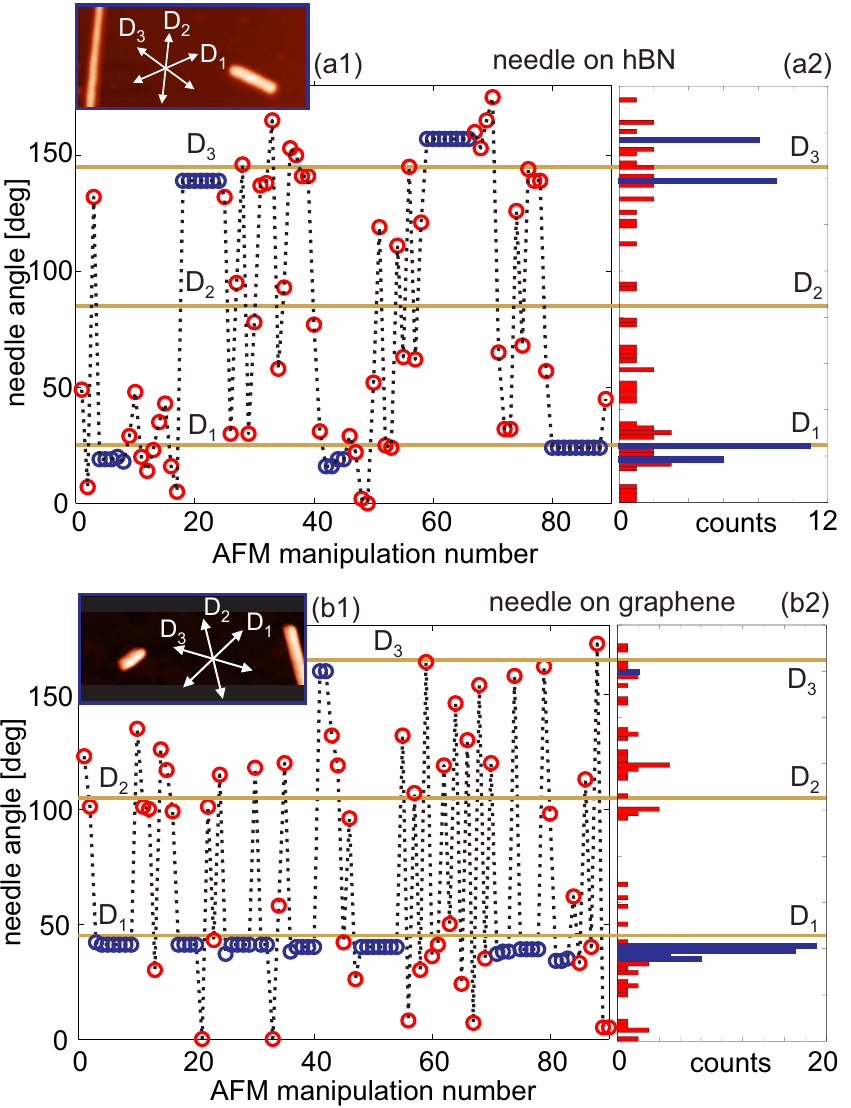}}
\caption{Change of the needle angle during AFM manipulations on (a1) hBN and (b1) Gr. The needle angle is defined as the angle between x-axis of the AFM scanner and long needle axis (LNA or needle direction). Parts (a2) and (b2) give the corresponding histograms for hBN and Gr, respectively. Horizontal solid lines $\mathrm{D_1-D_3}$ mark the preferential growth directions. $\mathrm{D_1-D_3}$ are also denoted in the insets with topographic images in parts (a1) and (b1). For the particular samples shown in the insets, D1-D3 directions denote LNA(R) for hBN, and LNA(L) for Gr. Histogram peaks (blue bars around $\mathrm{D_1}$ and $\mathrm{D_3}$) mark sequences where the needle fragments were translated along the registry states. The needle fragments are translated if the angle stays the same between two successive manipulation steps corresponding to red circles, while they are rotated if the angle changes between two successive manipulation steps. The point pairs where one point is below and other one above direction $\mathrm{D_2}$ correspond to the rotations across the registry state $\mathrm{D_2}$. }
\label{angle_distribution}
\end{figure}

The evolution and the histogram of the needle angle (calculated with respect to the x-axis for all manipulation steps) are presented in Figs. \ref{angle_distribution}(a1) and \ref{angle_distribution}(a2) for hBN substrate, and in Figs. \ref{angle_distribution}(b1) and \ref{angle_distribution}(b2) for Gr substrate. In Figs. \ref{angle_distribution}(a1) and \ref{angle_distribution}(b1), the arrays of successive points with the same needle angle denote the needle translations. Therefore, the needle on hBN was translated along direction $\mathrm{D_1}$ for steps 4-8, 42-45, and 80-88, and along $\mathrm{D_3}$ for steps 18-24 and 59-66. Directions $\mathrm{D_1}-\mathrm{D_3}$ mark the preferential growth directions as depicted in the inset of Fig. \ref{angle_distribution}(a) with the AFM topography image. They were found according to the growth directions of two long adjacent 6P needles and the six-fold symmetry of the hBN substrate (more details are provided in the description of Fig. S4 of ESI). For the Gr substrate, the needle was translated along direction $\mathrm{D_1}$ for steps 3-9, 17-20, 25-29, 36-40, 48-54, 71-73, 75-78, and 81-83, whereas translations along $\mathrm{D_3}$ were rarely observed, only in the two steps 41-42. Similar to the previous case, three preferential growth directions were marked with $\mathrm{D_1}-\mathrm{D_3}$ in the inset of Fig. \ref{angle_distribution}(b) with AFM topography image. They were determined according to the position of the adjacent long needle and the six-fold symmetry of Gr. In Figs. \ref{angle_distribution}(a2) and \ref{angle_distribution}(b2), the corresponding histograms of the needle angle are presented. The peaks in the histograms are clearly located around the preferential growth directions.

According to these results, we identified preferential directions for the sliding of 6P needles on hBN and Gr. These directions match quite well the preferential growth directions of the needles on both substrates, and they will be called registry states in the following. Although they are closely related to the commensurate contact planes between two crystal lattices, we believe that this is a more proper term, because only "flat-lying" 6P molecules in the bottom needle layer are commensurate with Gr and hBN. The registry states can be imagined as rails which define needle trajectories. Needles just slide along these rails, \textit{i.e.}, registry states, although pushed in a different direction.

During AFM manipulations, besides translations, we observed needle rotations across the registry states. They correspond to pairs of points in Figs. \ref{angle_distribution}(a1) and \ref{angle_distribution}(b1), with one point above and the second one below the line for $\mathrm{D_2}$. The sliding along direction $\mathrm{D_2}$ was not observed, neither for Gr nor for hBN because the angle between $\mathrm{D_2}$ and the manipulation direction is close to $90^\circ$. As a result, the applied torque was always too large leading to needle rotations across the registry state defined by $\mathrm{D_2}$. By measuring lateral forces during needle rotations, it was possible to map the existing friction anisotropy of the underlying substrates. This will be analyzed in detail in the next section.

\subsection{Friction anisotropy}

\begin{figure}
\centerline{\includegraphics[width=8.3cm]{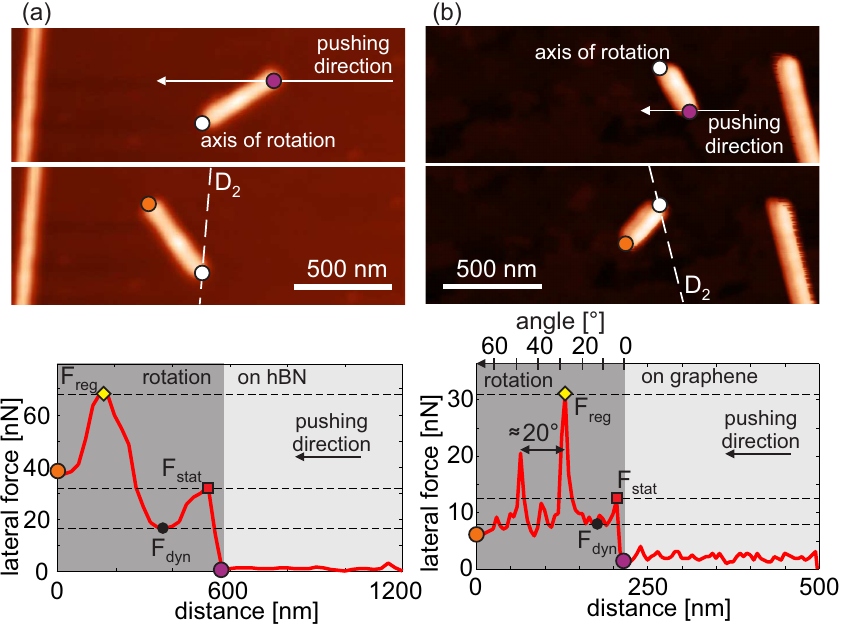}}
\caption{Rotations of 6P needle fragments across the registry state determined by direction $\mathrm{D_2}$: (a) on hBN ($z$ scale is 15~nm), (b) on Gr ($z$ scale is 10~nm). Top row: topographic images before AFM manipulation. Middle row: topographic images after the AFM manipulations. Bottom row: force profiles during the AFM manipulations. Arrows mark pushing directions and the path of pushing. Dashed lines denote the registry state $\mathrm{D_2}$.}
\label{force_images1}
\end{figure}

Typical images for the rotations on hBN and Gr substrates are presented in Figs. \ref{force_images1}(a) and \ref{force_images1}(b), respectively. Topographic images before and after the rotation are shown in the top and middle row, respectively, whereas the corresponding lateral force profile during AFM probe movement is given in the bottom row. As can be seen, first the AFM tip slides on the bare 2D material substrate, while the lateral force and thus the corresponding friction are low. Then, the AFM tip approaches the end of a needle fragment (purple dot) and starts to push the needle. This initial movement is described with an increase of the lateral force to the level $F_\mathrm{stat}$ (red square) which corresponds to the static friction \cite{Schirmeisen_friction_duality, Schirmeisen_scaling_laws, Baykara_nat_comm, Schirmeisen_jap_2007}. The needle is out of the registry at the beginning of the rotation, so the resulting friction between the needle and underlying substrate is low. For this reason, the lateral force drops from $F_\mathrm{stat}$ to $F_\mathrm{dyn}$ (black circle) corresponding to dynamic friction \cite{Schirmeisen_friction_duality, Schirmeisen_scaling_laws, Baykara_nat_comm, Schirmeisen_jap_2007}. With further rotation, the needle falls into the registry determined by direction $\mathrm{D_2}$, accompanied by a significant increase of the lateral force to $F_\mathrm{reg}$ (yellow diamond). After crossing the registry, the lateral force drops down (orange circle).

\begin{figure}
\centerline{\includegraphics[width=8.3cm]{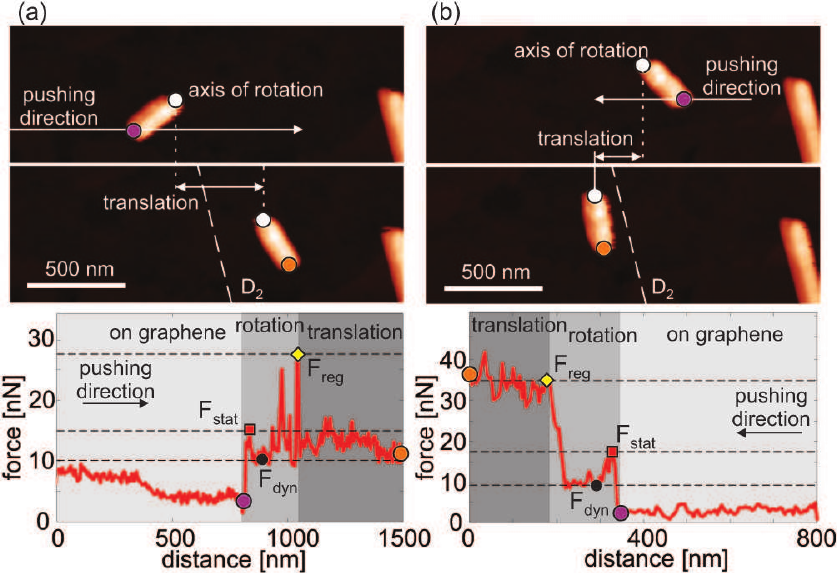}}
\caption{Rotation on Gr together with translation: (a) the needle is out of the registry state (defined by direction $\mathrm{D_2}$) during the translation, and (b) the needle remains in the registry state during the translation. Top row: topographic images before AFM manipulation. Middle row: topographic images after the AFM manipulations. Bottom row: lateral force profiles during the AFM manipulations. $z$ scale in the images is 10~nm. Arrows mark pushing directions and the path of pushing. Dashed lines denote the registry state $\mathrm{D_2}$.}
\label{force_images2}
\end{figure}

Figure \ref{force_images2} presents cases on Gr, where the needles are rotated across a registry state and simultaneously also translated, as can be seen by comparing to a reference point in the image, \textit{i.e.}, the end of a long as-grown needle. Figure \ref{force_images2}(a) demonstrates a case where the needle fragment is out of the registry state during the translation. In the force profile, again there are three already mentioned levels, namely, static friction at the beginning, dynamic friction after the needle is moved, and then a significant increase of the force when the needle is crossing the registry state defined by direction $\mathrm{D_2}$. After the needle passes across the registry state, the lateral force fluctuates between $F_\mathrm{stat}$ and $F_\mathrm{dyn}$. In this region, the needle is sliding on the Gr substrate, but is not falling into a registry state. On the other hand, in the example presented in Fig. \ref{force_images2}(b), after reaching of the high level $F_\mathrm{reg}$, the force practically stays on the same level until the end of moving. In this case, the needle is aligned in direction $\mathrm{D_2}$ at the end of the movement, meaning that after it felt into the registry, it remains in this state during the further translation.

\begin{figure}
\centerline{\includegraphics[width=7cm]{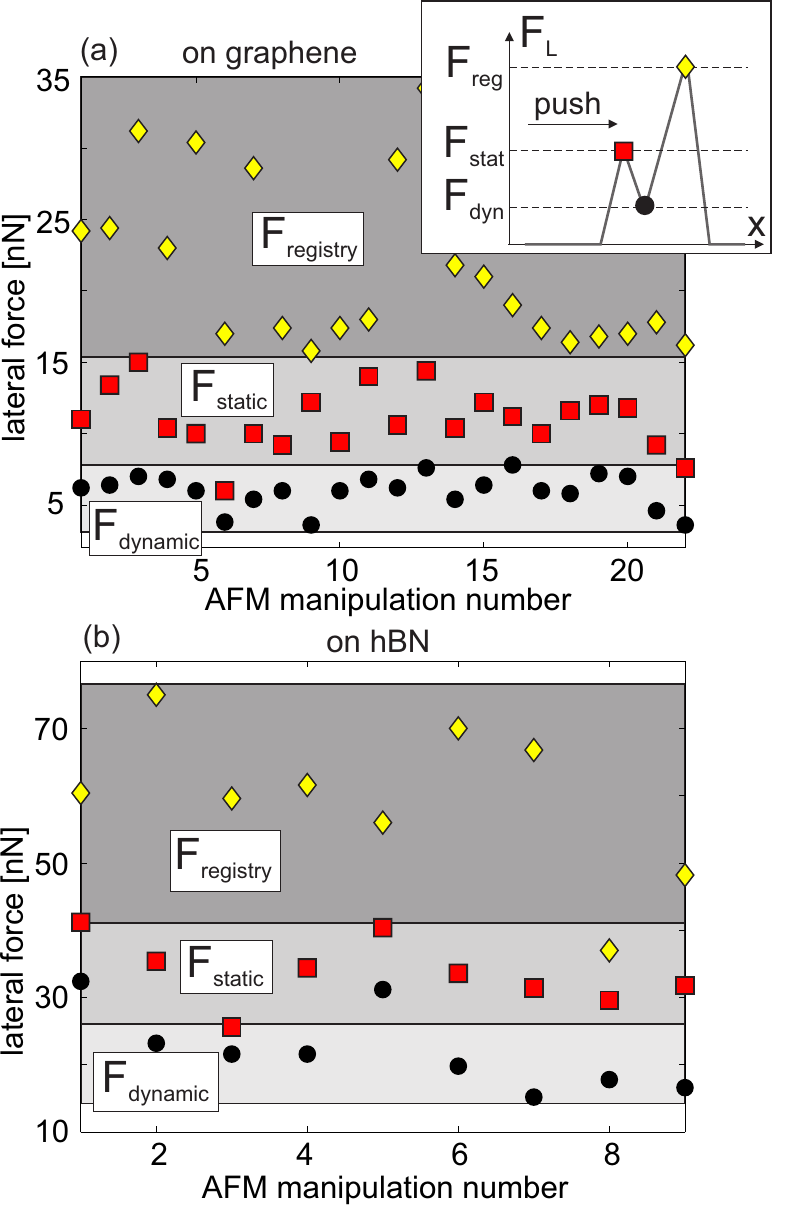}}
\caption{Characteristic lateral force ($F_\mathrm{L}$) levels $F_\mathrm{stat}$, $F_\mathrm{dyn}$, and $F_\mathrm{reg}$ during needle rotations on (a) Gr and (b) hBN.}
\label{force_distribution}
\end{figure}

Distributions of the characteristic force levels $F_\mathrm{stat}$, $F_\mathrm{dyn}$, and $F_\mathrm{reg}$ during all recorded needle rotations are presented in Figs. \ref{force_distribution}(a) and \ref{force_distribution}(b) for the manipulations on Gr and hBN, respectively. The characteristic force levels are very well distributed into three distinct ranges corresponding to static and dynamic friction, and as well as the friction in the registry state. As can be seen, $F_\mathrm{reg}$ is approximately 5 or 3 times higher than $F_\mathrm{dyn}$ on Gr and hBN, respectively, clearly indicating a significant friction anisotropy. Besides the described scenarios for needle rotations, we observed also cases where the needles were initially positioned in registry states. Then, the lateral force started from $F_\mathrm{reg}$ at the beginning of the rotation and then dropped. During some rotations, the registry state was not achieved at all due to a too small rotation angle. Since we could not measure all three force levels of interest in these cases, such cases were excluded from the analysis.

Now we return to a speciality only observed for the rotation of 6P needle fragments on Gr. In both Figs. \ref{force_images1}(b) and \ref{force_images2}(a), two peaks in the lateral force are observed during the rotation across the registry state. The case with a pure rotation (without translation) was given in Fig. \ref{force_images1}(b). Here, it was possible to approximately transform a distance into an angle according to the initial and final angles between the needle and the x-axis (the angle axis is indicated in the top of the force profile in Fig. \ref{force_images1}(b)). As can be seen, two peaks are separated by around $20^\circ$ from each other. Other images for the rotations on Gr together with lateral force profiles are provided in ESI in Fig. S3.  Figures S3(c), S3(l), and S3(p) present cases of pure rotations where the angle between two peaks was always observed to be around $20^\circ$. All other cases in Fig. S3 contain combined manipulations, consisting of both rotations and translations. For this reason, it was not possible to transform a distance into an angle. Still, all lateral force profiles in Fig. S3 as well as in Fig. \ref{force_images2}(a) exhibit such double peaks during needle manipulations. On the other hand, in the case of hBN, always only single peaks in the lateral force were observed as can be seen in Fig. \ref{force_images1}(a).

\subsection{MD simulations of needle movement}

\begin{figure}
\centerline{\includegraphics[width=6cm]{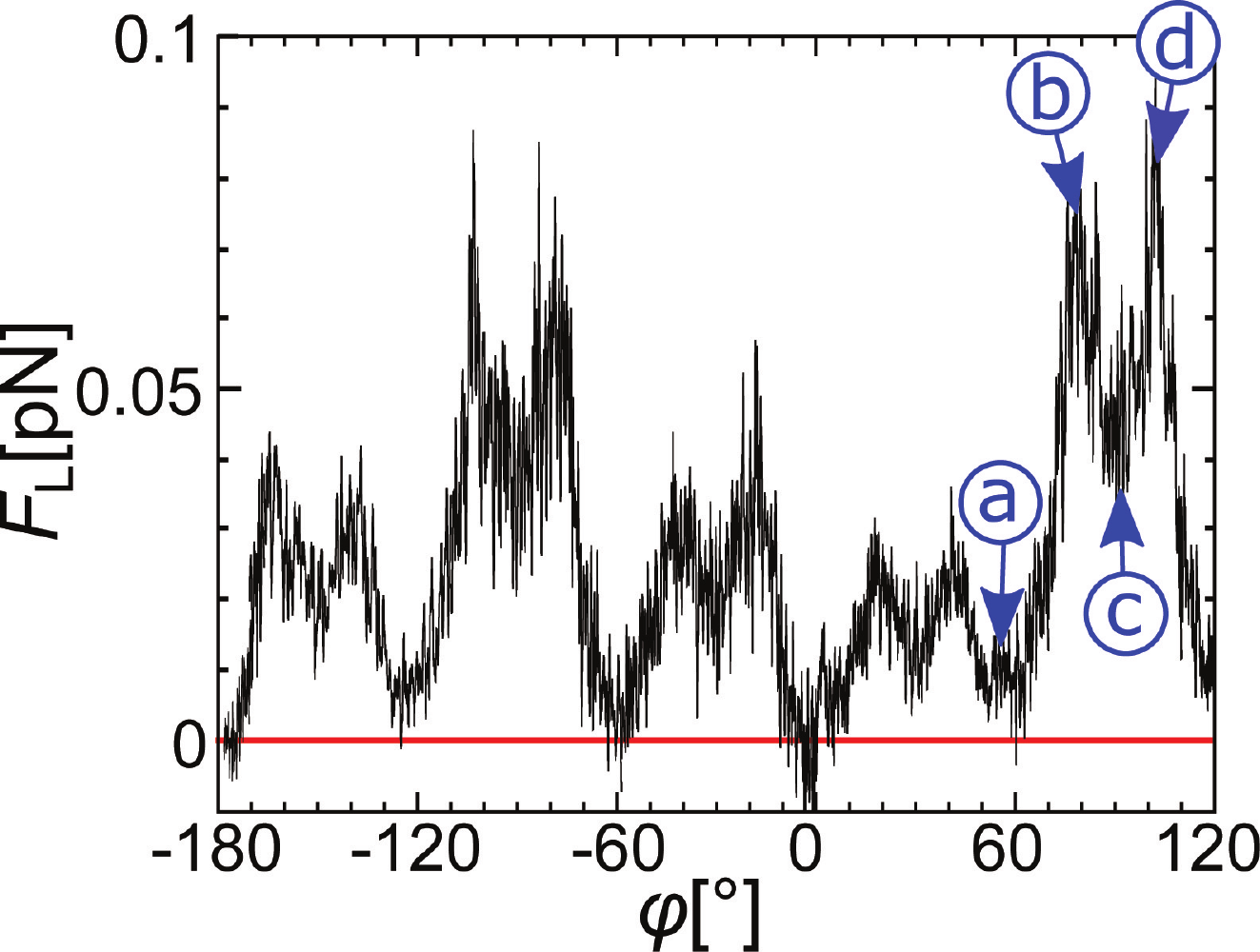}}
\caption{MD simulation results for the change of the lateral force with rotation of a needle fragment composed of $64 \times 4 \times 4$ 6P molecules. $F_{\rm L}$ is the mean lateral force of the bare Gr substrate acting on the needle. The results are presented for both clockwise (negative angles $\phi$) and anticlockwise rotation (positive angles $\phi$). Configurations for typical points (a), (b), (c), and (d) are indicated in Fig. \ref{md_configurations}.}
\label{md_friction_force}
\end{figure}

\begin{figure}
\centerline{\includegraphics[width=8.3cm]{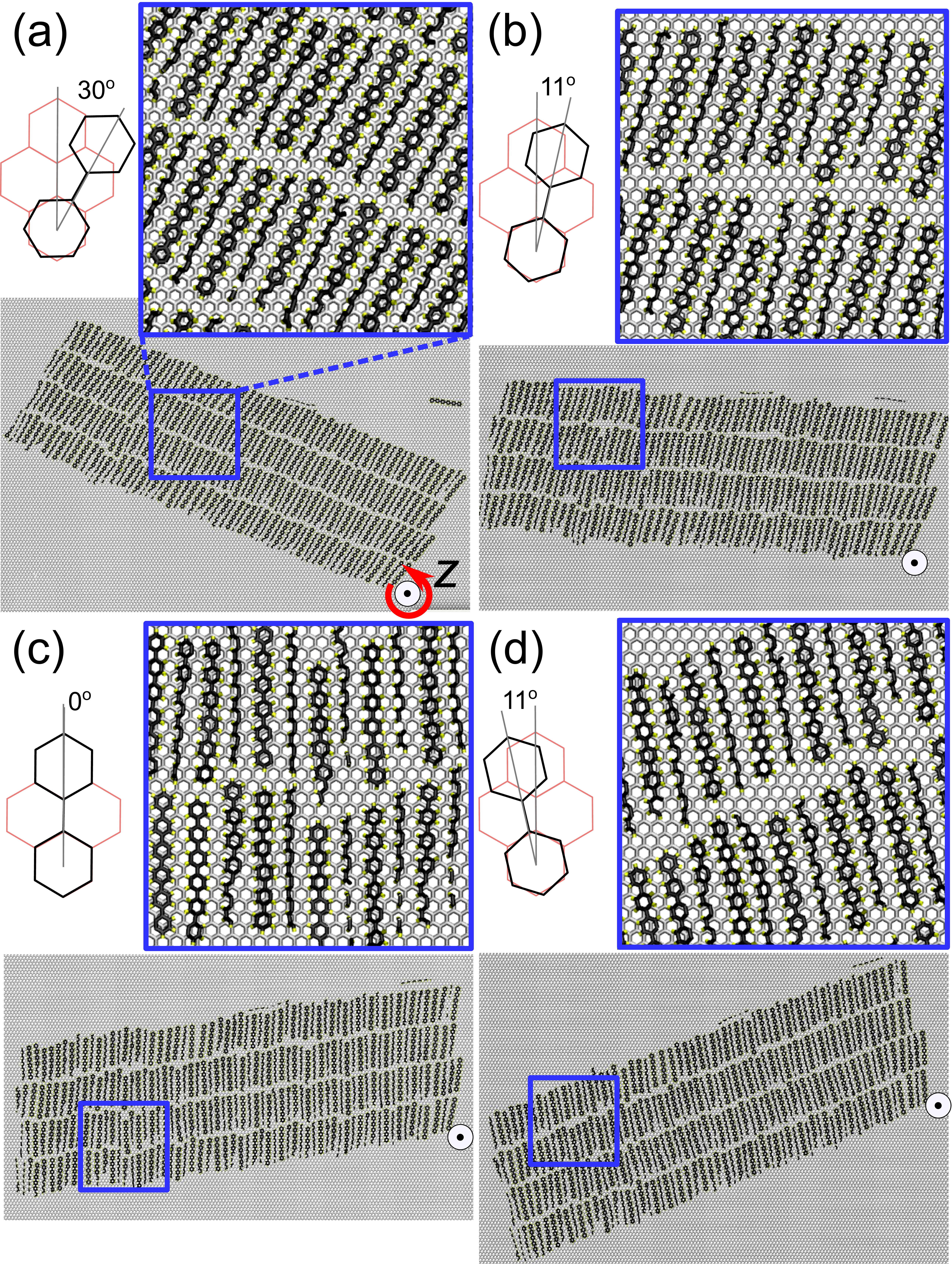}}
\caption{Snapshots of the bottom layer of a 6P needle on Gr obtained by MD simulations during needle rotations, shown at 4 typical stages during the rotation: (a) global minimum of the lateral force when 6P molecules are $30^\circ$ away from the Gr armchair direction, \textit{i.e.} aligned with the Gr zigzag direction, (b) first maximum of the lateral force when 6P molecules are $11^\circ$ away from the Gr armchair direction in the clockwise direction, (c) local minimum of the lateral force when 6P molecules are aligned with the Gr armchair direction, (d) second maximum of the the lateral force when 6P molecules are $11^\circ$ away from the Gr armchair direction in anticlockwise direction. The corresponding zooms of domains within the blue squares and a schematic representation of relative orientation between Gr and 6P molecule are presented. The Gr lattice is indicated in red and two phenyl rings of the 6P molecules are shown in black. The Gr armchair direction is oriented along the y-axis. The red arrow denotes the rotation direction (counterclockwise).}
\label{md_configurations}
\end{figure}

The results of MD simulations for the determination of the lateral force during 6P needle rotations (both clockwise and anticlockwise) on Gr are shown in Fig. \ref{md_friction_force} as a function of rotation angle $\phi$. The orientations of the needle and 6P molecules with respect to Gr at four characteristic points (a)-(d) (indicated in Fig. \ref{md_friction_force}) are depicted in Fig. \ref{md_configurations}. The animation of the needle rotation is given in ESI (supplementary movie 3). As can be seen from Fig. \ref{md_friction_force}, the friction force is approximately a periodic function of the rotation angle, with a period of about $60^\circ$, because of the six fold symmetry of Gr. Every period contains two peaks at characteristic points (b) and (d) with increased lateral force. The angular separation between these two peaks is in all periods around $20^\circ$.

As can be seen from the configurations in Figs. \ref{md_configurations}(b) and \ref{md_configurations}(d), at points (b) and (d), the long axis of 6P molecules is $11^\circ$ away from the Gr armchair direction (aligned along y-axis). Thus, at points (b) and (d), the LMA directions are rotationally commensurate with the substrate \cite{Teichert_nanolett}. Therefore, MD simulations indicate two close registry states, tilted by $\pm$11$^{\circ}$ from an armchair direction of Gr either in clockwise or anticlockwise direction. When 6P molecules are aligned with the Gr armchair direction, there is a local minimum in the lateral force at point (c). The global minimum in the lateral force is reached at point (a), when 6P molecules are aligned with the Gr zigzag direction.

As explained in Fig. \ref{structure_fig}, there are not only three, but three pairs of preferential growth directions. They are denoted with LNA, while two directions within a single LNA pair are marked with L and R (chiral pairs), and they are separated for Gr by around $10^\circ$ as schematically displayed in Fig. \ref{structure_fig}(g). Still, only one direction, either L or R, in each pair can be a true registry state for the same short needle. In this state, both rotational and translational epitaxial relations between a "flat-lying" 6P molecule and the Gr lattice are conserved.

As mentioned earlier, 6P molecules that are in contact with the Gr have their preferential adsorption site with the LMA tilted by $\sim$11$^{\circ}$ from an armchair direction \cite{Teichert_nanolett, balzer2013Gr6Porientation}. Two chiral pairs, L and R, are then separated by $\sim$22$^{\circ}$. During a needle rotation, it is possible that the needle (LNA direction) falls in a state where the molecules in contact with Gr are only rotationally commensurate with the substrate, but do not match the exact positions as would be the case for the true commensurate state and for as-grown needles. This situation is depicted in Fig. \ref{structure_fig}(g) for the needle with a true commensurate state denoted with LNA(L), and when it is rotated by $22^\circ$ in the clockwise direction (then it is aligned with the direction marked with LNA(L)$-22^\circ$). Such states should still present sufficiently deep potential energy minima for the "flat-lying" molecules at the interface with Gr. This fact really explains the existence of the two friction maxima (commensurate states) during the rotation of the 6P needle on Gr which are separated by around $20^\circ$ as confirmed by both experiments and MD simulations.

In the case of hBN, 6P molecules in face-on position have their LMA oriented exactly parallel to the armchair direction of hBN \cite{Matkovic_scirep_2016}. Therefore, only one friction maximum appears when the LMA of 6P molecule is rotated across the armchair direction of hBN, which is in accordance with the experimental results in Fig. \ref{force_images1}(a).

\subsection{Preferential sliding directions}

\begin{figure}
\centerline{\includegraphics[width=8.3cm]{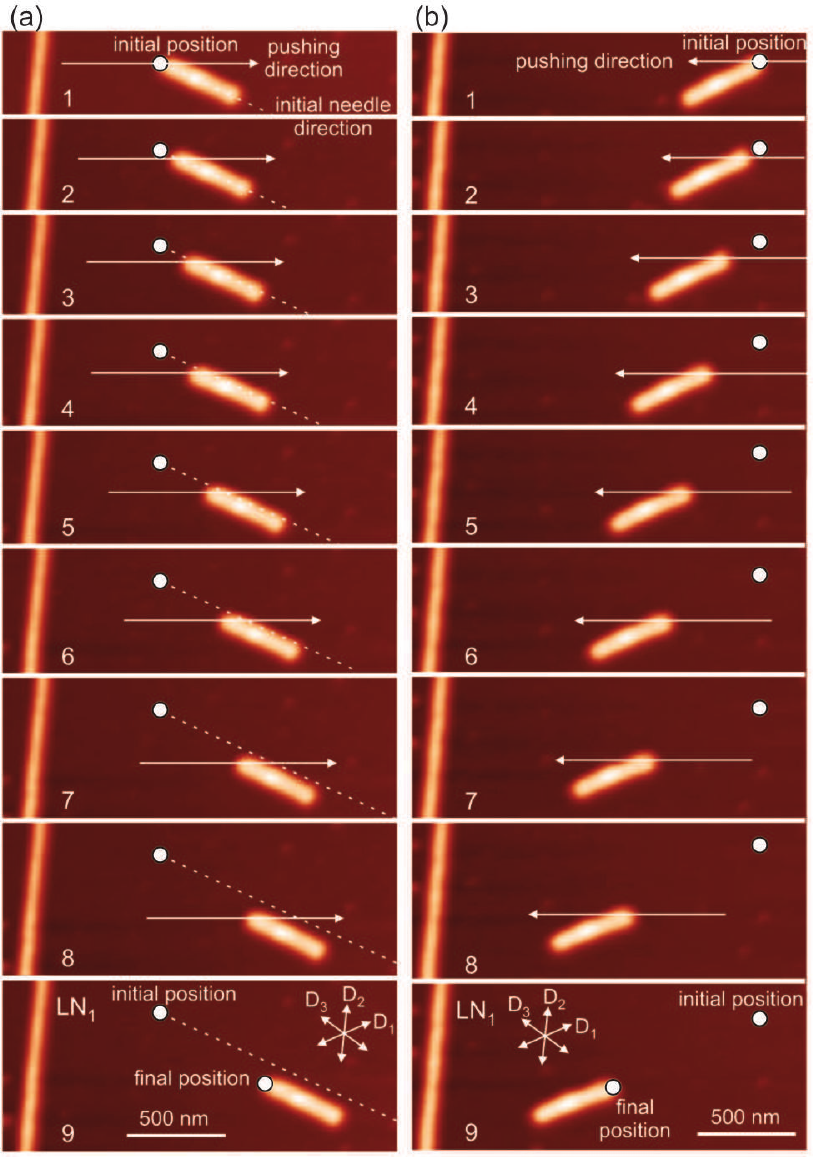}}
\caption{Sequences of AFM images for 6P needle sliding on hBN: (a) sliding along preferred direction $\mathrm{D_3}$ and (b) $\mathrm{D_1}$. The long needle $\mathrm{LN_1}$ is taken as a reference. Arrows mark pushing directions and the path of pushing. Dotted lines in column (a) denote the initial needle direction making visible a small needle shift in x-direction as well, not only along the preferential direction $\mathrm{D_3}$. $z$ scale in all images is 15~nm.}
\label{translation_hbn}
\end{figure}

The observed friction anisotropy also explains the existence of preferential sliding directions where short needles are just translated along the registry states. The results for the translation of a short needle on hBN are shown in Fig. \ref{translation_hbn}. It represents two sequences of 9 needle positions during pushing. The part of the long needle $\mathrm{LN_1}$ on the left side of the images was taken as a reference object. As can be seen, the short needle was pushed along the x-axis from its left and right ending, while it was translated along the directions $\mathrm{D_3}$ (Fig. \ref{translation_hbn}(a)) and $\mathrm{D_1}$ (Fig. \ref{translation_hbn}(b)), respectively. The resulting shifts along these directions were below $100 \ \mathrm{nm}$, and have been determined by the distance along which the AFM tip was in contact with the needle.

\begin{figure}
\centerline{\includegraphics[width=8.3cm]{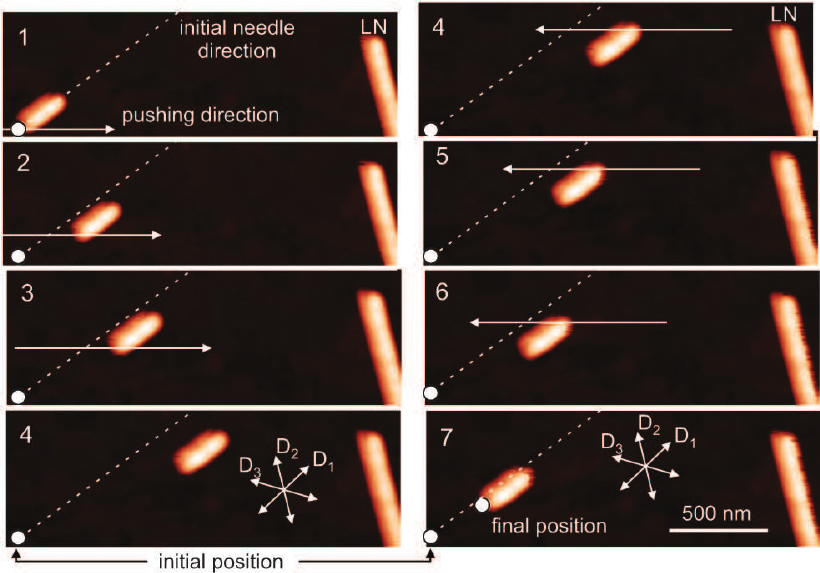}}
\caption{Sequence of AFM images for 6P needle sliding on Gr. The end of the long needle (LN) is taken as a reference point. Arrows mark pushing directions and the path of pushing. Oblique dashed lines denote the initial needle direction making visible a small needle shift in x-direction as well, not only along the preferential direction $\mathrm{D_1}$. $z$ scale in the images is 10~nm.}
\label{translation_gr}
\end{figure}

A characteristic example for the preferential sliding on Gr is presented in Fig. \ref{translation_gr}. Here, the end of a long needle LN on the right side serves as a reference object. The short needle was pushed both in positive (steps 1-4, left hand side of Fig. \ref{translation_gr}) and negative x direction (steps 4-7, right hand side of Fig. \ref{translation_gr}). Still, as a result of this pushing, the needle was just translated along the preferential direction $\mathrm{D_1}$.

As can be seen in Fig. \ref{angle_distribution}, experimentally measured preferential sliding directions slightly differ from the marked preferential growth directions $\mathrm{D_1}-\mathrm{D_3}$. There are several possible reasons for these deviations. Preferential directions $\mathrm{D_1}-\mathrm{D_3}$ were determined from directions of adjacent long needles (two of them in the case of hBN and one needle in the case of Gr) and the six-fold symmetry of both substrates. This could lead to a small error of a few degrees. 6P needles could also be slightly rotated from the preferential growth directions during AFM manipulations. For example, on the hBN substrate, 6P molecules prefer to be oriented exactly along armchair directions. Small rotations of the molecules with respect to armchair directions by a few degrees lead only to a slight increase of the adsorption energy as shown in Ref. \cite{Matkovic_scirep_2016}. Still, even such states can be regarded as commensurate for 6P molecules, and can define preferential sliding directions.

Oblique dashed lines in Figs. \ref{translation_hbn}(a) and \ref{translation_gr} denote the initial needle direction. As can be seen, during the sliding, needles are not just moved along the preferential directions, but they could be slightly shifted to an adjacent registry state. Inspite of this shift, they still stay aligned with the preferential sliding directions. Therefore, Gr and hBN substrates can be imagined as arrays of parallel rails. When pushed by the AFM probe, 6P needles slide along a single rail, but at some points, they can jump to the next parallel rail due to the pushing force. After this jumping, the sliding continues along the same preferential direction. Slight shifts to adjacent registry states can be explained in the following way. Direction of the registry state is the principal direction of friction. If the needle slides along the principal direction, the friction force is parallel, but with the opposite direction. However, if the needle is not completely in the registry state (for example, misaligned by several degrees), or if the pushing force slightly moves it from the registry state, then an additional force component appears along the direction normal to the registry state \cite{Campione2012,Campione2013,Bennewitz_prb_2014}, and this additional force can be responsible for the observed needle movement in the lateral direction (with respect to the direction of the registry state).

\section{Conclusions}
To summarize, using combined AFM based manipulation and MD simulations, we investigated the influence of the epitaxial relations between organic 6P needles and Gr/hBN substrates on the resulting needle movement and the underlying friction. It was demonstrated that the preferential growth directions, split by $\pm$5$^\circ$ from high symmetry directions of Gr and hBN, determine registry states for short 6P needle fragments that have been cut by AFM manipulations out of long needles. During the AFM manipulations of short 6P needles, we observed both, their translations and rotations. In the case of the translations, we revealed that the preferential sliding directions coincide with the preferential growth directions of a crystallite with a particular chirality, and that these directions are in accordance with the underlying epitaxial relations. In the case of rotations across registry states, the friction was increased by around 5 and 3 times on Gr and hBN respectively, compared to the dynamic friction out of the registry. Therefore, our results reveal that the organic nanocrystallites behave on 2D materials as if they would follow invisible rails of commensurate directions, and tend to slide along or switch between these "rails". These results provide new insights into frictional properties of 2D materials and also prove that AFM manipulation of nanoparticles is an efficient technique to study friction in vdW heterostructures.

\section*{Conflicts of interest}
There are no conflicts to declare.

\section*{Acknowledgements}
This work is supported by the Serbian Ministry of Education, Science and Technological Development under Projects No. OI171005, OI171017, and 451-03-01039/2015-09/40, by Austrian Science Fund (FWF) through project I~1788-N20, by Austrian Academic Exchange Services through the project SRB 09/2016, and in part by COST Action MP1303. A. Matkovi\'c acknowledges the support from the Lise Meitner fellowship by Austrian Science Fund (FWF): M~2323-N36. Numerical simulations were run on the PARADOX supercomputing facility at the Scientific Computing Laboratory of the Institute of Physics Belgrade.






\begin{mcitethebibliography}{72}
\providecommand*{\natexlab}[1]{#1}
\providecommand*{\mciteSetBstSublistMode}[1]{}
\providecommand*{\mciteSetBstMaxWidthForm}[2]{}
\providecommand*{\mciteBstWouldAddEndPuncttrue}
  {\def\EndOfBibitem{\unskip.}}
\providecommand*{\mciteBstWouldAddEndPunctfalse}
  {\let\EndOfBibitem\relax}
\providecommand*{\mciteSetBstMidEndSepPunct}[3]{}
\providecommand*{\mciteSetBstSublistLabelBeginEnd}[3]{}
\providecommand*{\EndOfBibitem}{}
\mciteSetBstSublistMode{f}
\mciteSetBstMaxWidthForm{subitem}
{(\emph{\alph{mcitesubitemcount}})}
\mciteSetBstSublistLabelBeginEnd{\mcitemaxwidthsubitemform\space}
{\relax}{\relax}

\bibitem[Donnet and Erdemir(2004)]{Erdemir_review}
C.~Donnet and A.~Erdemir, \emph{Surf. Coat. Technol.}, 2004, \textbf{180}, 76
  -- 84\relax
\mciteBstWouldAddEndPuncttrue
\mciteSetBstMidEndSepPunct{\mcitedefaultmidpunct}
{\mcitedefaultendpunct}{\mcitedefaultseppunct}\relax
\EndOfBibitem
\bibitem[Lee \emph{et~al.}(2010)Lee, Li, Kalb, Liu, Berger, Carpick, and
  Hone]{friction_Carpick}
C.~Lee, Q.~Li, W.~Kalb, X.-Z. Liu, H.~Berger, R.~W. Carpick and J.~Hone,
  \emph{Science}, 2010, \textbf{328}, 76--80\relax
\mciteBstWouldAddEndPuncttrue
\mciteSetBstMidEndSepPunct{\mcitedefaultmidpunct}
{\mcitedefaultendpunct}{\mcitedefaultseppunct}\relax
\EndOfBibitem
\bibitem[Berman \emph{et~al.}(2014)Berman, Erdemir, and
  Sumant]{graphene_lubricant}
D.~Berman, A.~Erdemir and A.~V. Sumant, \emph{Mater. Today}, 2014, \textbf{17},
  31--42\relax
\mciteBstWouldAddEndPuncttrue
\mciteSetBstMidEndSepPunct{\mcitedefaultmidpunct}
{\mcitedefaultendpunct}{\mcitedefaultseppunct}\relax
\EndOfBibitem
\bibitem[Filleter \emph{et~al.}(2009)Filleter, McChesney, Bostwick, Rotenberg,
  Emtsev, Seyller, Horn, and Bennewitz]{friction_Bennewitz}
T.~Filleter, J.~L. McChesney, A.~Bostwick, E.~Rotenberg, K.~V. Emtsev,
  T.~Seyller, K.~Horn and R.~Bennewitz, \emph{Phys. Rev. Lett.}, 2009,
  \textbf{102}, 086102\relax
\mciteBstWouldAddEndPuncttrue
\mciteSetBstMidEndSepPunct{\mcitedefaultmidpunct}
{\mcitedefaultendpunct}{\mcitedefaultseppunct}\relax
\EndOfBibitem
\bibitem[Kim \emph{et~al.}(2011)Kim, Lee, Lee, Lee, Jang, Ahn, Kim, and
  Lee]{Lee_ACSnano}
K.-S. Kim, H.-J. Lee, C.~Lee, S.-K. Lee, H.~Jang, J.-H. Ahn, J.-H. Kim and
  H.-J. Lee, \emph{ACS Nano}, 2011, \textbf{5}, 5107--5114\relax
\mciteBstWouldAddEndPuncttrue
\mciteSetBstMidEndSepPunct{\mcitedefaultmidpunct}
{\mcitedefaultendpunct}{\mcitedefaultseppunct}\relax
\EndOfBibitem
\bibitem[Klemenz \emph{et~al.}(2014)Klemenz, Pastewka, Balakrishna, Caron,
  Bennewitz, and Moseler]{friction_reduction_Moseler}
A.~Klemenz, L.~Pastewka, S.~G. Balakrishna, A.~Caron, R.~Bennewitz and
  M.~Moseler, \emph{Nano Lett.}, 2014, \textbf{14}, 7145--7152\relax
\mciteBstWouldAddEndPuncttrue
\mciteSetBstMidEndSepPunct{\mcitedefaultmidpunct}
{\mcitedefaultendpunct}{\mcitedefaultseppunct}\relax
\EndOfBibitem
\bibitem[Berman \emph{et~al.}(2014)Berman, Deshmukh, Sankaranarayanan, Erdemir,
  and Sumant]{Berman_adv_func_mat}
D.~Berman, S.~A. Deshmukh, S.~K. R.~S. Sankaranarayanan, A.~Erdemir and A.~V.
  Sumant, \emph{Adv. Funct. Mater.}, 2014, \textbf{24}, 6640--6646\relax
\mciteBstWouldAddEndPuncttrue
\mciteSetBstMidEndSepPunct{\mcitedefaultmidpunct}
{\mcitedefaultendpunct}{\mcitedefaultseppunct}\relax
\EndOfBibitem
\bibitem[Vasi\'c \emph{et~al.}(2017)Vasi\'c, Matkovi\'c, Ralevi\'c, Beli\'c,
  and Gaji\'c]{Vasic_carbon_2017}
B.~Vasi\'c, A.~Matkovi\'c, U.~Ralevi\'c, M.~Beli\'c and R.~Gaji\'c,
  \emph{Carbon}, 2017, \textbf{120}, 137 -- 144\relax
\mciteBstWouldAddEndPuncttrue
\mciteSetBstMidEndSepPunct{\mcitedefaultmidpunct}
{\mcitedefaultendpunct}{\mcitedefaultseppunct}\relax
\EndOfBibitem
\bibitem[Sheehan and Lieber(1996)]{Sheehan_science_1996}
P.~E. Sheehan and C.~M. Lieber, \emph{Science}, 1996, \textbf{272},
  1158--1161\relax
\mciteBstWouldAddEndPuncttrue
\mciteSetBstMidEndSepPunct{\mcitedefaultmidpunct}
{\mcitedefaultendpunct}{\mcitedefaultseppunct}\relax
\EndOfBibitem
\bibitem[Falvo \emph{et~al.}(2000)Falvo, Steele, Taylor, and
  Superfine]{Falvo_prb_2000}
M.~R. Falvo, J.~Steele, R.~M. Taylor and R.~Superfine, \emph{Phys. Rev. B},
  2000, \textbf{62}, R10665--R10667\relax
\mciteBstWouldAddEndPuncttrue
\mciteSetBstMidEndSepPunct{\mcitedefaultmidpunct}
{\mcitedefaultendpunct}{\mcitedefaultseppunct}\relax
\EndOfBibitem
\bibitem[Tranvouez \emph{et~al.}(2009)Tranvouez, Orieux, Boer-Duchemin,
  Devillers, Huc, Comtet, and Dujardin]{Tranvouez_nanotechnology_2009}
E.~Tranvouez, A.~Orieux, E.~Boer-Duchemin, C.~H. Devillers, V.~Huc, G.~Comtet
  and G.~Dujardin, \emph{Nanotechnology}, 2009, \textbf{20}, 165304\relax
\mciteBstWouldAddEndPuncttrue
\mciteSetBstMidEndSepPunct{\mcitedefaultmidpunct}
{\mcitedefaultendpunct}{\mcitedefaultseppunct}\relax
\EndOfBibitem
\bibitem[Balakrishna \emph{et~al.}(2014)Balakrishna, de~Wijn, and
  Bennewitz]{Bennewitz_prb_2014}
S.~G. Balakrishna, A.~S. de~Wijn and R.~Bennewitz, \emph{Phys. Rev. B}, 2014,
  \textbf{89}, 245440\relax
\mciteBstWouldAddEndPuncttrue
\mciteSetBstMidEndSepPunct{\mcitedefaultmidpunct}
{\mcitedefaultendpunct}{\mcitedefaultseppunct}\relax
\EndOfBibitem
\bibitem[Sheehan and Lieber(2017)]{Sheehan_nanolett}
P.~E. Sheehan and C.~M. Lieber, \emph{Nano Lett.}, 2017, \textbf{17},
  4116--4121\relax
\mciteBstWouldAddEndPuncttrue
\mciteSetBstMidEndSepPunct{\mcitedefaultmidpunct}
{\mcitedefaultendpunct}{\mcitedefaultseppunct}\relax
\EndOfBibitem
\bibitem[Dienwiebel \emph{et~al.}(2004)Dienwiebel, Verhoeven, Pradeep, Frenken,
  Heimberg, and Zandbergen]{Dienwiebel_prl_2004}
M.~Dienwiebel, G.~S. Verhoeven, N.~Pradeep, J.~W.~M. Frenken, J.~A. Heimberg
  and H.~W. Zandbergen, \emph{Phys. Rev. Lett.}, 2004, \textbf{92},
  126101\relax
\mciteBstWouldAddEndPuncttrue
\mciteSetBstMidEndSepPunct{\mcitedefaultmidpunct}
{\mcitedefaultendpunct}{\mcitedefaultseppunct}\relax
\EndOfBibitem
\bibitem[Dirk \emph{et~al.}(2008)Dirk, Claudia, M\"onninghoff, Fuchs,
  Schirmeisen, and Schwarz]{Schirmeisen_friction_duality}
D.~Dirk, R.~Claudia, T.~M\"onninghoff, H.~Fuchs, A.~Schirmeisen and U.~D.
  Schwarz, \emph{Phys. Rev. Lett.}, 2008, \textbf{101}, 125505\relax
\mciteBstWouldAddEndPuncttrue
\mciteSetBstMidEndSepPunct{\mcitedefaultmidpunct}
{\mcitedefaultendpunct}{\mcitedefaultseppunct}\relax
\EndOfBibitem
\bibitem[Dietzel \emph{et~al.}(2013)Dietzel, Feldmann, Schwarz, Fuchs, and
  Schirmeisen]{Schirmeisen_scaling_laws}
D.~Dietzel, M.~Feldmann, U.~D. Schwarz, H.~Fuchs and A.~Schirmeisen,
  \emph{Phys. Rev. Lett.}, 2013, \textbf{111}, 235502\relax
\mciteBstWouldAddEndPuncttrue
\mciteSetBstMidEndSepPunct{\mcitedefaultmidpunct}
{\mcitedefaultendpunct}{\mcitedefaultseppunct}\relax
\EndOfBibitem
\bibitem[Koren \emph{et~al.}(2015)Koren, L{\"o}rtscher, Rawlings, Knoll, and
  Duerig]{duerig_science_2015}
E.~Koren, E.~L{\"o}rtscher, C.~Rawlings, A.~W. Knoll and U.~Duerig,
  \emph{Science}, 2015, \textbf{348}, 679--683\relax
\mciteBstWouldAddEndPuncttrue
\mciteSetBstMidEndSepPunct{\mcitedefaultmidpunct}
{\mcitedefaultendpunct}{\mcitedefaultseppunct}\relax
\EndOfBibitem
\bibitem[Kawai \emph{et~al.}(2016)Kawai, Benassi, Gnecco, S\"ode, Pawlak, Feng,
  M\"ullen, Passerone, Pignedoli, Ruffieux, Fasel, and
  Meyer]{Meyer_science_2016}
S.~Kawai, A.~Benassi, E.~Gnecco, H.~S\"ode, R.~Pawlak, X.~Feng, K.~M\"ullen,
  D.~Passerone, C.~A. Pignedoli, P.~Ruffieux, R.~Fasel and E.~Meyer,
  \emph{Science}, 2016, \textbf{351}, 957\relax
\mciteBstWouldAddEndPuncttrue
\mciteSetBstMidEndSepPunct{\mcitedefaultmidpunct}
{\mcitedefaultendpunct}{\mcitedefaultseppunct}\relax
\EndOfBibitem
\bibitem[Cihan \emph{et~al.}(2016)Cihan, \.Ipek, Durgun, and
  Baykara]{Baykara_nat_comm}
E.~Cihan, S.~\.Ipek, E.~Durgun and M.~Z. Baykara, \emph{Nat. Comm.}, 2016,
  \textbf{7}, 12055\relax
\mciteBstWouldAddEndPuncttrue
\mciteSetBstMidEndSepPunct{\mcitedefaultmidpunct}
{\mcitedefaultendpunct}{\mcitedefaultseppunct}\relax
\EndOfBibitem
\bibitem[Hirano \emph{et~al.}(1991)Hirano, Shinjo, Kaneko, and
  Murata]{Hirano_Shinjo_prl_1991}
M.~Hirano, K.~Shinjo, R.~Kaneko and Y.~Murata, \emph{Phys. Rev. Lett.}, 1991,
  \textbf{67}, 2642--2645\relax
\mciteBstWouldAddEndPuncttrue
\mciteSetBstMidEndSepPunct{\mcitedefaultmidpunct}
{\mcitedefaultendpunct}{\mcitedefaultseppunct}\relax
\EndOfBibitem
\bibitem[Liu \emph{et~al.}(2012)Liu, Yang, Grey, Liu, Liu, Wang, Yang, Cheng,
  and Zheng]{prl_microscale_superlubricity}
Z.~Liu, J.~Yang, F.~Grey, J.~Z. Liu, Y.~Liu, Y.~Wang, Y.~Yang, Y.~Cheng and
  Q.~Zheng, \emph{Phys. Rev. Lett.}, 2012, \textbf{108}, 205503\relax
\mciteBstWouldAddEndPuncttrue
\mciteSetBstMidEndSepPunct{\mcitedefaultmidpunct}
{\mcitedefaultendpunct}{\mcitedefaultseppunct}\relax
\EndOfBibitem
\bibitem[de~Wijn(2012)]{deWijn_prb_2012}
A.~S. de~Wijn, \emph{Phys. Rev. B}, 2012, \textbf{86}, 085429\relax
\mciteBstWouldAddEndPuncttrue
\mciteSetBstMidEndSepPunct{\mcitedefaultmidpunct}
{\mcitedefaultendpunct}{\mcitedefaultseppunct}\relax
\EndOfBibitem
\bibitem[Feng \emph{et~al.}(2013)Feng, Kwon, Park, and
  Salmeron]{Salmeron_acs_nano}
X.~Feng, S.~Kwon, J.~Y. Park and M.~Salmeron, \emph{ACS Nano}, 2013,
  \textbf{7}, 1718--1724\relax
\mciteBstWouldAddEndPuncttrue
\mciteSetBstMidEndSepPunct{\mcitedefaultmidpunct}
{\mcitedefaultendpunct}{\mcitedefaultseppunct}\relax
\EndOfBibitem
\bibitem[Liu \emph{et~al.}(2014)Liu, Grey, and Zheng]{zheng_sci_rep_2014}
Y.~Liu, F.~Grey and Q.~Zheng, \emph{Sci. Rep.}, 2014, \textbf{4}, 4875\relax
\mciteBstWouldAddEndPuncttrue
\mciteSetBstMidEndSepPunct{\mcitedefaultmidpunct}
{\mcitedefaultendpunct}{\mcitedefaultseppunct}\relax
\EndOfBibitem
\bibitem[Berman \emph{et~al.}(2015)Berman, Deshmukh, Sankaranarayanan, Erdemir,
  and Sumant]{berman_science_2015}
D.~Berman, S.~A. Deshmukh, S.~K. R.~S. Sankaranarayanan, A.~Erdemir and A.~V.
  Sumant, \emph{Science}, 2015, \textbf{348}, 1118--1122\relax
\mciteBstWouldAddEndPuncttrue
\mciteSetBstMidEndSepPunct{\mcitedefaultmidpunct}
{\mcitedefaultendpunct}{\mcitedefaultseppunct}\relax
\EndOfBibitem
\bibitem[Lee \emph{et~al.}(2009)Lee, Lee, Seo, Eom, and Lee]{friction_Lee}
H.~Lee, N.~Lee, Y.~Seo, J.~Eom and S.~Lee, \emph{Nanotechnology}, 2009,
  \textbf{20}, 325701\relax
\mciteBstWouldAddEndPuncttrue
\mciteSetBstMidEndSepPunct{\mcitedefaultmidpunct}
{\mcitedefaultendpunct}{\mcitedefaultseppunct}\relax
\EndOfBibitem
\bibitem[Kwon \emph{et~al.}(2012)Kwon, Ko, Jeon, Kim, and Park]{friction_Park}
S.~Kwon, J.-H. Ko, K.-J. Jeon, Y.-H. Kim and J.~Y. Park, \emph{Nano Lett.},
  2012, \textbf{12}, 6043--6048\relax
\mciteBstWouldAddEndPuncttrue
\mciteSetBstMidEndSepPunct{\mcitedefaultmidpunct}
{\mcitedefaultendpunct}{\mcitedefaultseppunct}\relax
\EndOfBibitem
\bibitem[Fessler \emph{et~al.}(2014)Fessler, Eren, Gysin, Glatzel, and
  Meyer]{friction_Meyer}
G.~Fessler, B.~Eren, U.~Gysin, T.~Glatzel and E.~Meyer, \emph{Appl. Phys.
  Lett.}, 2014, \textbf{104}, 041910\relax
\mciteBstWouldAddEndPuncttrue
\mciteSetBstMidEndSepPunct{\mcitedefaultmidpunct}
{\mcitedefaultendpunct}{\mcitedefaultseppunct}\relax
\EndOfBibitem
\bibitem[Chen and Filleter(2015)]{Filleter_GO}
H.~Chen and T.~Filleter, \emph{Nanotechnology}, 2015, \textbf{26}, 135702\relax
\mciteBstWouldAddEndPuncttrue
\mciteSetBstMidEndSepPunct{\mcitedefaultmidpunct}
{\mcitedefaultendpunct}{\mcitedefaultseppunct}\relax
\EndOfBibitem
\bibitem[Dietzel \emph{et~al.}(2007)Dietzel, M\"onninghoff, Jansen, Fuchs,
  Ritter, Schwarz, and Schirmeisen]{Schirmeisen_jap_2007}
D.~Dietzel, T.~M\"onninghoff, L.~Jansen, H.~Fuchs, C.~Ritter, U.~D. Schwarz and
  A.~Schirmeisen, \emph{J. Appl. Phys.}, 2007, \textbf{102}, 084306\relax
\mciteBstWouldAddEndPuncttrue
\mciteSetBstMidEndSepPunct{\mcitedefaultmidpunct}
{\mcitedefaultendpunct}{\mcitedefaultseppunct}\relax
\EndOfBibitem
\bibitem[Koma(1999)]{Koma}
A.~Koma, \emph{J. Cryst. Growth}, 1999, \textbf{201}, 236 -- 241\relax
\mciteBstWouldAddEndPuncttrue
\mciteSetBstMidEndSepPunct{\mcitedefaultmidpunct}
{\mcitedefaultendpunct}{\mcitedefaultseppunct}\relax
\EndOfBibitem
\bibitem[Kratzer and Teichert(2016)]{Kratzer_review}
M.~Kratzer and C.~Teichert, \emph{Nanotechnology}, 2016, \textbf{27},
  292001\relax
\mciteBstWouldAddEndPuncttrue
\mciteSetBstMidEndSepPunct{\mcitedefaultmidpunct}
{\mcitedefaultendpunct}{\mcitedefaultseppunct}\relax
\EndOfBibitem
\bibitem[Hlawacek \emph{et~al.}(2011)Hlawacek, Khokhar, van Gastel, Poelsema,
  and Teichert]{Teichert_nanolett}
G.~Hlawacek, F.~S. Khokhar, R.~van Gastel, B.~Poelsema and C.~Teichert,
  \emph{Nano Lett.}, 2011, \textbf{11}, 333--337\relax
\mciteBstWouldAddEndPuncttrue
\mciteSetBstMidEndSepPunct{\mcitedefaultmidpunct}
{\mcitedefaultendpunct}{\mcitedefaultseppunct}\relax
\EndOfBibitem
\bibitem[Matkovi\'c \emph{et~al.}(2016)Matkovi\'c, Genser, L\"uftner, Kratzer,
  Gaji\'c, Puschnig, and Teichert]{Matkovic_scirep_2016}
A.~Matkovi\'c, J.~Genser, D.~L\"uftner, M.~Kratzer, R.~Gaji\'c, P.~Puschnig and
  C.~Teichert, \emph{Sci. Rep.}, 2016, \textbf{6}, 38519\relax
\mciteBstWouldAddEndPuncttrue
\mciteSetBstMidEndSepPunct{\mcitedefaultmidpunct}
{\mcitedefaultendpunct}{\mcitedefaultseppunct}\relax
\EndOfBibitem
\bibitem[Lee \emph{et~al.}(2014)Lee, Lee, van~der Zande, Han, Cui, Arefe,
  Nuckolls, Heinz, Hone, and Kim]{Kim_apl_mater}
G.-H. Lee, C.-H. Lee, A.~M. van~der Zande, M.~Han, X.~Cui, G.~Arefe,
  C.~Nuckolls, T.~F. Heinz, J.~Hone and P.~Kim, \emph{APL Mater.}, 2014,
  \textbf{2}, 092511\relax
\mciteBstWouldAddEndPuncttrue
\mciteSetBstMidEndSepPunct{\mcitedefaultmidpunct}
{\mcitedefaultendpunct}{\mcitedefaultseppunct}\relax
\EndOfBibitem
\bibitem[Lee \emph{et~al.}(2014)Lee, Schiros, Santos, Kim, Yager, Kang, Lee,
  Yu, Watanabe, Taniguchi, Hone, Kaxiras, Nuckolls, and Kim]{Kim_adv_mater}
C.-H. Lee, T.~Schiros, E.~J.~G. Santos, B.~Kim, K.~G. Yager, S.~J. Kang,
  S.~Lee, J.~Yu, K.~Watanabe, T.~Taniguchi, J.~Hone, E.~Kaxiras, C.~Nuckolls
  and P.~Kim, \emph{Adv. Mater.}, 2014, \textbf{26}, 2812--2817\relax
\mciteBstWouldAddEndPuncttrue
\mciteSetBstMidEndSepPunct{\mcitedefaultmidpunct}
{\mcitedefaultendpunct}{\mcitedefaultseppunct}\relax
\EndOfBibitem
\bibitem[Zhang \emph{et~al.}(2016)Zhang, Qiao, Gao, Hu, He, Wu, Yang, Xu, Li,
  Shi, Ji, Wang, Wang, Xiao, Xu, Xu, and Wang]{Wang_prl}
Y.~Zhang, J.~Qiao, S.~Gao, F.~Hu, D.~He, B.~Wu, Z.~Yang, B.~Xu, Y.~Li, Y.~Shi,
  W.~Ji, P.~Wang, X.~Wang, M.~Xiao, H.~Xu, J.-B. Xu and X.~Wang, \emph{Phys.
  Rev. Lett.}, 2016, \textbf{116}, 016602\relax
\mciteBstWouldAddEndPuncttrue
\mciteSetBstMidEndSepPunct{\mcitedefaultmidpunct}
{\mcitedefaultendpunct}{\mcitedefaultseppunct}\relax
\EndOfBibitem
\bibitem[Jariwala \emph{et~al.}(2016)Jariwala, Howell, Chen, Kang, Sangwan,
  Filippone, Turrisi, Marks, Lauhon, and Hersam]{Hersam_nanolett}
D.~Jariwala, S.~L. Howell, K.-S. Chen, J.~Kang, V.~K. Sangwan, S.~A. Filippone,
  R.~Turrisi, T.~J. Marks, L.~J. Lauhon and M.~C. Hersam, \emph{Nano Lett.},
  2016, \textbf{16}, 497--503\relax
\mciteBstWouldAddEndPuncttrue
\mciteSetBstMidEndSepPunct{\mcitedefaultmidpunct}
{\mcitedefaultendpunct}{\mcitedefaultseppunct}\relax
\EndOfBibitem
\bibitem[He \emph{et~al.}(1999)He, M\"user, and Robbins]{Muser_science_1999}
G.~He, M.~H. M\"user and M.~O. Robbins, \emph{Science}, 1999, \textbf{284},
  1650--1652\relax
\mciteBstWouldAddEndPuncttrue
\mciteSetBstMidEndSepPunct{\mcitedefaultmidpunct}
{\mcitedefaultendpunct}{\mcitedefaultseppunct}\relax
\EndOfBibitem
\bibitem[Dietzel \emph{et~al.}(2017)Dietzel, Brndiar, {\v S}tich, and
  Schirmeisen]{Schirmeisen_chemical_interactions}
D.~Dietzel, J.~Brndiar, I.~{\v S}tich and A.~Schirmeisen, \emph{ACS Nano},
  2017, \textbf{11}, 7642--7647\relax
\mciteBstWouldAddEndPuncttrue
\mciteSetBstMidEndSepPunct{\mcitedefaultmidpunct}
{\mcitedefaultendpunct}{\mcitedefaultseppunct}\relax
\EndOfBibitem
\bibitem[Overney \emph{et~al.}(1994)Overney, Takano, Fujihira, Paulus, and
  Ringsdorf]{fr_anisotropy_organic_prl_1994}
R.~M. Overney, H.~Takano, M.~Fujihira, W.~Paulus and H.~Ringsdorf, \emph{Phys.
  Rev. Lett.}, 1994, \textbf{72}, 3546--3549\relax
\mciteBstWouldAddEndPuncttrue
\mciteSetBstMidEndSepPunct{\mcitedefaultmidpunct}
{\mcitedefaultendpunct}{\mcitedefaultseppunct}\relax
\EndOfBibitem
\bibitem[Carpick \emph{et~al.}(1999)Carpick, Sasaki, and
  Burns]{carpick_trib_lett_1999}
R.~W. Carpick, D.~Y. Sasaki and A.~R. Burns, \emph{Tribol. Lett.}, 1999,
  \textbf{7}, 79--85\relax
\mciteBstWouldAddEndPuncttrue
\mciteSetBstMidEndSepPunct{\mcitedefaultmidpunct}
{\mcitedefaultendpunct}{\mcitedefaultseppunct}\relax
\EndOfBibitem
\bibitem[Kalihari \emph{et~al.}(2010)Kalihari, Haugstad, and
  Frisbie]{fr_anisotropy_organic_prl_2010a}
V.~Kalihari, G.~Haugstad and C.~D. Frisbie, \emph{Phys. Rev. Lett.}, 2010,
  \textbf{104}, 086102\relax
\mciteBstWouldAddEndPuncttrue
\mciteSetBstMidEndSepPunct{\mcitedefaultmidpunct}
{\mcitedefaultendpunct}{\mcitedefaultseppunct}\relax
\EndOfBibitem
\bibitem[Campione and Fumagalli(2010)]{fr_anisotropy_organic_prl_2010b}
M.~Campione and E.~Fumagalli, \emph{Phys. Rev. Lett.}, 2010, \textbf{105},
  166103\relax
\mciteBstWouldAddEndPuncttrue
\mciteSetBstMidEndSepPunct{\mcitedefaultmidpunct}
{\mcitedefaultendpunct}{\mcitedefaultseppunct}\relax
\EndOfBibitem
\bibitem[Perez-Rodriguez \emph{et~al.}(2017)Perez-Rodriguez, Barrena,
  Fernandez, Gnecco, and Ocal]{ocal_nanoscale2017}
A.~Perez-Rodriguez, E.~Barrena, A.~Fernandez, E.~Gnecco and C.~Ocal,
  \emph{Nanoscale}, 2017, \textbf{9}, 5589--5596\relax
\mciteBstWouldAddEndPuncttrue
\mciteSetBstMidEndSepPunct{\mcitedefaultmidpunct}
{\mcitedefaultendpunct}{\mcitedefaultseppunct}\relax
\EndOfBibitem
\bibitem[Novoselov \emph{et~al.}(2004)Novoselov, Geim, Morozov, Jiang, Zhang,
  Dubonos, , Grigorieva, and Firsov]{novoselov_science2004}
K.~S. Novoselov, A.~K. Geim, S.~V. Morozov, D.~Jiang, Y.~Zhang, S.~V. Dubonos,
  , I.~V. Grigorieva and A.~A. Firsov, \emph{Science}, 2004, \textbf{306},
  666--669\relax
\mciteBstWouldAddEndPuncttrue
\mciteSetBstMidEndSepPunct{\mcitedefaultmidpunct}
{\mcitedefaultendpunct}{\mcitedefaultseppunct}\relax
\EndOfBibitem
\bibitem[Lopez-Otero(1978)]{lopez1978hot}
A.~Lopez-Otero, \emph{Thin Solid Films}, 1978, \textbf{49}, 3--57\relax
\mciteBstWouldAddEndPuncttrue
\mciteSetBstMidEndSepPunct{\mcitedefaultmidpunct}
{\mcitedefaultendpunct}{\mcitedefaultseppunct}\relax
\EndOfBibitem
\bibitem[Potocar \emph{et~al.}(2011)Potocar, Lorbek, Nabok, Shen, Tumbek,
  Hlawacek, Puschnig, Ambrosch-Draxl, Teichert, and Winkler]{potocar_prb}
T.~Potocar, S.~Lorbek, D.~Nabok, Q.~Shen, L.~Tumbek, G.~Hlawacek, P.~Puschnig,
  C.~Ambrosch-Draxl, C.~Teichert and A.~Winkler, \emph{Phys. Rev. B}, 2011,
  \textbf{83}, 075423\relax
\mciteBstWouldAddEndPuncttrue
\mciteSetBstMidEndSepPunct{\mcitedefaultmidpunct}
{\mcitedefaultendpunct}{\mcitedefaultseppunct}\relax
\EndOfBibitem
\bibitem[Balzer \emph{et~al.}(2013)Balzer, Henrichsen, Klarskov, Booth, Sun,
  Parisi, Schiek, and B{\o}ggild]{balzer2013Gr6Porientation}
F.~Balzer, H.~H. Henrichsen, M.~B. Klarskov, T.~J. Booth, R.~Sun, J.~Parisi,
  M.~Schiek and P.~B{\o}ggild, \emph{Nanotechnology}, 2013, \textbf{25},
  035602\relax
\mciteBstWouldAddEndPuncttrue
\mciteSetBstMidEndSepPunct{\mcitedefaultmidpunct}
{\mcitedefaultendpunct}{\mcitedefaultseppunct}\relax
\EndOfBibitem
\bibitem[Kratzer \emph{et~al.}(2013)Kratzer, Klima, Teichert, Vasi{\'c},
  Matkovi{\'c}, Ralevi{\'c}, and Gaji{\'c}]{kratzer2013temperature}
M.~Kratzer, S.~Klima, C.~Teichert, B.~Vasi{\'c}, A.~Matkovi{\'c},
  U.~Ralevi{\'c} and R.~Gaji{\'c}, \emph{J. Vac. Sci. Technol. B}, 2013,
  \textbf{31}, 04D114\relax
\mciteBstWouldAddEndPuncttrue
\mciteSetBstMidEndSepPunct{\mcitedefaultmidpunct}
{\mcitedefaultendpunct}{\mcitedefaultseppunct}\relax
\EndOfBibitem
\bibitem[Hlawacek and Teichert(2013)]{hlawacek2013nucleation}
G.~Hlawacek and C.~Teichert, \emph{J. Phys. Condens. Matter}, 2013,
  \textbf{25}, 143202\relax
\mciteBstWouldAddEndPuncttrue
\mciteSetBstMidEndSepPunct{\mcitedefaultmidpunct}
{\mcitedefaultendpunct}{\mcitedefaultseppunct}\relax
\EndOfBibitem
\bibitem[Simbrunner(2013)]{simbrunner2013epitaxial}
C.~Simbrunner, \emph{Semicond. Sci. Technol.}, 2013, \textbf{28}, 053001\relax
\mciteBstWouldAddEndPuncttrue
\mciteSetBstMidEndSepPunct{\mcitedefaultmidpunct}
{\mcitedefaultendpunct}{\mcitedefaultseppunct}\relax
\EndOfBibitem
\bibitem[Baker \emph{et~al.}(1993)Baker, Fratini, Resch, Knachel, Adams, Socci,
  and Farmer]{baker1993crystal}
K.~N. Baker, A.~V. Fratini, T.~Resch, H.~C. Knachel, W.~W. Adams, E.~P. Socci
  and B.~L. Farmer, \emph{Polymer}, 1993, \textbf{34}, 1571--1587\relax
\mciteBstWouldAddEndPuncttrue
\mciteSetBstMidEndSepPunct{\mcitedefaultmidpunct}
{\mcitedefaultendpunct}{\mcitedefaultseppunct}\relax
\EndOfBibitem
\bibitem[Hutter and Bechhoefer(1993)]{Hutter_calibration}
J.~L. Hutter and J.~Bechhoefer, \emph{Rev. Sci. Instrum.}, 1993, \textbf{64},
  1868\relax
\mciteBstWouldAddEndPuncttrue
\mciteSetBstMidEndSepPunct{\mcitedefaultmidpunct}
{\mcitedefaultendpunct}{\mcitedefaultseppunct}\relax
\EndOfBibitem
\bibitem[Kjelstrup-Hansen \emph{et~al.}(2006)Kjelstrup-Hansen, Hansen, Rubahn,
  and B{\o}ggild]{Bogild_small_2006}
J.~Kjelstrup-Hansen, O.~Hansen, H.-G. Rubahn and P.~B{\o}ggild, \emph{Small},
  2006, \textbf{2}, 660--666\relax
\mciteBstWouldAddEndPuncttrue
\mciteSetBstMidEndSepPunct{\mcitedefaultmidpunct}
{\mcitedefaultendpunct}{\mcitedefaultseppunct}\relax
\EndOfBibitem
\bibitem[Junno \emph{et~al.}(1995)Junno, Deppert, Montelius, and
  Samuelson]{Samuelson_APL}
T.~Junno, K.~Deppert, L.~Montelius and L.~Samuelson, \emph{Appl. Phys. Lett.},
  1995, \textbf{66}, 3627--3629\relax
\mciteBstWouldAddEndPuncttrue
\mciteSetBstMidEndSepPunct{\mcitedefaultmidpunct}
{\mcitedefaultendpunct}{\mcitedefaultseppunct}\relax
\EndOfBibitem
\bibitem[Theil~Hansen \emph{et~al.}(1998)Theil~Hansen, K{\"u}hle, S{\o}rensen,
  Bohr, and Lindelof]{Lindelof_Nanotechnology}
L.~Theil~Hansen, A.~K{\"u}hle, A.~H. S{\o}rensen, J.~Bohr and P.~E. Lindelof,
  \emph{Nanotechnology}, 1998, \textbf{9}, 337\relax
\mciteBstWouldAddEndPuncttrue
\mciteSetBstMidEndSepPunct{\mcitedefaultmidpunct}
{\mcitedefaultendpunct}{\mcitedefaultseppunct}\relax
\EndOfBibitem
\bibitem[Gnecco \emph{et~al.}(2010)Gnecco, Rao, Mougin, Chandrasekar, and
  Meyer]{Gnecco2010}
E.~Gnecco, A.~Rao, K.~Mougin, G.~Chandrasekar and E.~Meyer,
  \emph{Nanotechnology}, 2010, \textbf{21}, 215702\relax
\mciteBstWouldAddEndPuncttrue
\mciteSetBstMidEndSepPunct{\mcitedefaultmidpunct}
{\mcitedefaultendpunct}{\mcitedefaultseppunct}\relax
\EndOfBibitem
\bibitem[Varenberg \emph{et~al.}(2003)Varenberg, Etsion, and
  Halperin]{Varenberg}
M.~Varenberg, I.~Etsion and G.~Halperin, \emph{Rev. Sci. Instrum.}, 2003,
  \textbf{74}, 3362\relax
\mciteBstWouldAddEndPuncttrue
\mciteSetBstMidEndSepPunct{\mcitedefaultmidpunct}
{\mcitedefaultendpunct}{\mcitedefaultseppunct}\relax
\EndOfBibitem
\bibitem[Tersoff(1989)]{Tersoff_CC}
J.~Tersoff, \emph{Phys. Rev. B}, 1989, \textbf{39}, 5566--5568\relax
\mciteBstWouldAddEndPuncttrue
\mciteSetBstMidEndSepPunct{\mcitedefaultmidpunct}
{\mcitedefaultendpunct}{\mcitedefaultseppunct}\relax
\EndOfBibitem
\bibitem[Brooks \emph{et~al.}(2009)Brooks, Brooks, Mackerell, Nilsson,
  Petrella, Roux, Won, Archontis, Bartels, Boresch, Caflisch, Caves, Cui,
  Dinner, Feig, Fischer, Gao, Hodoscek, Im, Kuczera, Lazaridis, Ma,
  Ovchinnikov, Paci, Pastor, Post, Pu, Schaefer, Tidor, Venable, Woodcock, Wu,
  Yang, York, and Karplus]{CHARMM}
B.~R. Brooks, C.~L. Brooks, A.~D. Mackerell, L.~Nilsson, R.~J. Petrella,
  B.~Roux, Y.~Won, G.~Archontis, C.~Bartels, S.~Boresch, A.~Caflisch, L.~Caves,
  Q.~Cui, A.~R. Dinner, M.~Feig, S.~Fischer, J.~Gao, M.~Hodoscek, W.~Im,
  K.~Kuczera, T.~Lazaridis, J.~Ma, V.~Ovchinnikov, E.~Paci, R.~W. Pastor, C.~B.
  Post, J.~Z. Pu, M.~Schaefer, B.~Tidor, R.~M. Venable, H.~L. Woodcock, X.~Wu,
  W.~Yang, D.~M. York and M.~Karplus, \emph{J. Comput. Chem.}, 2009,
  \textbf{30}, 1545--1614\relax
\mciteBstWouldAddEndPuncttrue
\mciteSetBstMidEndSepPunct{\mcitedefaultmidpunct}
{\mcitedefaultendpunct}{\mcitedefaultseppunct}\relax
\EndOfBibitem
\bibitem[Kolmogorov and Crespi(2005)]{PhysRevB.71.235415}
A.~N. Kolmogorov and V.~H. Crespi, \emph{Phys. Rev. B}, 2005, \textbf{71},
  235415\relax
\mciteBstWouldAddEndPuncttrue
\mciteSetBstMidEndSepPunct{\mcitedefaultmidpunct}
{\mcitedefaultendpunct}{\mcitedefaultseppunct}\relax
\EndOfBibitem
\bibitem[Plimpton(1995)]{LAMMPS}
S.~Plimpton, \emph{J. Comput. Phys.}, 1995, \textbf{117}, 1--19\relax
\mciteBstWouldAddEndPuncttrue
\mciteSetBstMidEndSepPunct{\mcitedefaultmidpunct}
{\mcitedefaultendpunct}{\mcitedefaultseppunct}\relax
\EndOfBibitem
\bibitem[Hooks \emph{et~al.}(2001)Hooks, Fritz, and Ward]{Hooks2001}
D.~E. Hooks, T.~Fritz and M.~D. Ward, \emph{Adv. Mater.}, 2001, \textbf{13},
  227--241\relax
\mciteBstWouldAddEndPuncttrue
\mciteSetBstMidEndSepPunct{\mcitedefaultmidpunct}
{\mcitedefaultendpunct}{\mcitedefaultseppunct}\relax
\EndOfBibitem
\bibitem[Mannsfeld \emph{et~al.}(2005)Mannsfeld, Leo, and Fritz]{Mannsfeld2005}
S.~C.~B. Mannsfeld, K.~Leo and T.~Fritz, \emph{Phys. Rev. Lett.}, 2005,
  \textbf{94}, 056104\relax
\mciteBstWouldAddEndPuncttrue
\mciteSetBstMidEndSepPunct{\mcitedefaultmidpunct}
{\mcitedefaultendpunct}{\mcitedefaultseppunct}\relax
\EndOfBibitem
\bibitem[Haber \emph{et~al.}(2008)Haber, Resel, Thierry, Campione, Sassella,
  and Moret]{Haber2008}
T.~Haber, R.~Resel, A.~Thierry, M.~Campione, A.~Sassella and M.~Moret,
  \emph{Physica E: Low Dimens. Syst. Nanostruct.}, 2008, \textbf{41}, 133 --
  137\relax
\mciteBstWouldAddEndPuncttrue
\mciteSetBstMidEndSepPunct{\mcitedefaultmidpunct}
{\mcitedefaultendpunct}{\mcitedefaultseppunct}\relax
\EndOfBibitem
\bibitem[Raimondo \emph{et~al.}(2011)Raimondo, Moret, Campione, Borghesi, and
  Sassella]{Raimondo2011}
L.~Raimondo, M.~Moret, M.~Campione, A.~Borghesi and A.~Sassella, \emph{J. Phys.
  Chem. C}, 2011, \textbf{115}, 5880--5885\relax
\mciteBstWouldAddEndPuncttrue
\mciteSetBstMidEndSepPunct{\mcitedefaultmidpunct}
{\mcitedefaultendpunct}{\mcitedefaultseppunct}\relax
\EndOfBibitem
\bibitem[Raimondo \emph{et~al.}(2013)Raimondo, Fumagalli, Moret, Campione,
  Borghesi, and Sassella]{Raimondo2013}
L.~Raimondo, E.~Fumagalli, M.~Moret, M.~Campione, A.~Borghesi and A.~Sassella,
  \emph{J. Phys. Chem. C}, 2013, \textbf{117}, 13981--13988\relax
\mciteBstWouldAddEndPuncttrue
\mciteSetBstMidEndSepPunct{\mcitedefaultmidpunct}
{\mcitedefaultendpunct}{\mcitedefaultseppunct}\relax
\EndOfBibitem
\bibitem[Campione \emph{et~al.}(2006)Campione, Sassella, Moret, Papagni,
  Trabattoni, Resel, Lengyel, Marcon, and Raos]{Campione2006}
M.~Campione, A.~Sassella, M.~Moret, A.~Papagni, S.~Trabattoni, R.~Resel,
  O.~Lengyel, V.~Marcon and G.~Raos, \emph{J. Am. Chem. Soc.}, 2006,
  \textbf{128}, 13378--13387\relax
\mciteBstWouldAddEndPuncttrue
\mciteSetBstMidEndSepPunct{\mcitedefaultmidpunct}
{\mcitedefaultendpunct}{\mcitedefaultseppunct}\relax
\EndOfBibitem
\bibitem[Conache \emph{et~al.}(2009)Conache, Gray, Ribayrol, Fr\"{o}berg,
  Samuelson, Pettersson, and Montelius]{Conache2009}
G.~Conache, S.~M. Gray, A.~Ribayrol, L.~E. Fr\"{o}berg, L.~Samuelson,
  H.~Pettersson and L.~Montelius, \emph{Small}, 2009, \textbf{5},
  203--207\relax
\mciteBstWouldAddEndPuncttrue
\mciteSetBstMidEndSepPunct{\mcitedefaultmidpunct}
{\mcitedefaultendpunct}{\mcitedefaultseppunct}\relax
\EndOfBibitem
\bibitem[Campione \emph{et~al.}(2012)Campione, Trabattoni, and
  Moret]{Campione2012}
M.~Campione, S.~Trabattoni and M.~Moret, \emph{Tribol. Lett.}, 2012,
  \textbf{45}, 219--224\relax
\mciteBstWouldAddEndPuncttrue
\mciteSetBstMidEndSepPunct{\mcitedefaultmidpunct}
{\mcitedefaultendpunct}{\mcitedefaultseppunct}\relax
\EndOfBibitem
\bibitem[Campione and Capitani(2013)]{Campione2013}
M.~Campione and G.~C. Capitani, \emph{Nat. Geosci.}, 2013, \textbf{6},
  847\relax
\mciteBstWouldAddEndPuncttrue
\mciteSetBstMidEndSepPunct{\mcitedefaultmidpunct}
{\mcitedefaultendpunct}{\mcitedefaultseppunct}\relax
\EndOfBibitem
\end{mcitethebibliography}
\bibliographystyle{rsc} 

\providecommand*{\mcitethebibliography}{\thebibliography}
\csname @ifundefined\endcsname{endmcitethebibliography}
{\let\endmcitethebibliography\endthebibliography}{}

\end{document}